\documentclass[a4paper,fleqn,usenatbib]{mnras}

\usepackage[T1]{fontenc}
\usepackage{ae,aecompl}
\usepackage{amsmath}	
\usepackage{arydshln}

\usepackage{newtxtext,newtxmath}

\usepackage{graphicx}	
\usepackage{amssymb}	
\usepackage{gensymb}
\usepackage{mathtools}
\usepackage{enumitem}
\usepackage{multirow}

 \DeclareMathOperator*{\argmax}{arg\,max}

\title{Detection of non-thermal X-ray emission in the lobes and jets of Cygnus A}

\author[M.N. de Vries]{
M. N. de Vries$^{1}$ \thanks{E-mail: m.n.devries@uva.nl (KTS)},
M. W. Wise$^{2,1}$, 
D. Huppenkothen$^{3}$, 
P. E. J. Nulsen$^{4}$,
B. Snios$^{4}$, 
\newauthor M. J. Hardcastle$^{5}$,
M. Birkinshaw$^{6}$,
D. M. Worrall$^{6}$,
R. T. Duffy$^{6},$ and
B. R. McNamara$^{7,8}$\\
$^{1}$ Astronomical Institute "Anton Pannekoek", University of
Amsterdam, Science Park 904, 1098 XH Amsterdam, The
Netherlands\\
$^{2}$  ASTRON, Netherlands Institute for Radio Astronomy, Postbus
2, 7990 AA, Dwingeloo, The Netherlands \\
$^{3}$ Department of Astronomy, University of Washington, 3910 15th Ave NE, Seattle, WA 98195 \\
$^{4}$ Harvard-Smithsonian Center for Astrophysics, 60 Garden Street, Cambridge, MA 02138 \\
$^{5}$ School of Physics, Astronomy and Mathematics, University of Hertfordshire, College Lane, Hatfield AL10 9AB, UK \\
$^{6}$ HH Wills Physics Laboratory, University of Bristol, Tyndall Avenue, Bristol, BS8 1TL, UK \\
$^{7}$ Department of Physics and Astronomy, University of Waterloo, 200 University Ave W, Waterloo, ON N2L 3G1, Canada  \\
$^{8}$ Perimeter Institute for Theoretical Physics, 31 Caroline St. North, Waterloo, ON N2L 2Y5, Canada 
}

\date{Accepted XXX. Received YYY; in original form ZZZ}

\pubyear{2017}

\begin{document}
\label{firstpage}
\pagerange{\pageref{firstpage}--\pageref{lastpage}}
\maketitle

\begin{abstract}
We present a spectral analysis of the lobes and X-ray jets of Cygnus A, using more than 2 Ms of $\textit{Chandra}$ observations. The X-ray jets are misaligned with the radio jets and significantly wider. We detect non-thermal emission components in both lobes and jets. For the eastern lobe and jet, we find 1 keV flux densities of $71_{-10}^{+10}$ nJy and $24_{-4}^{+4}$ nJy, and photon indices of $1.72_{-0.03}^{+0.03}$ and $1.64_{-0.04}^{+0.04}$ respectively. For the western lobe and jet, we find flux densities of $50_{-13}^{+12}$ nJy and $13_{-5}^{+5}$ nJy, and photon indices of  $1.97_{-0.10}^{+0.23}$ and $1.86_{-0.12}^{+0.18}$ respectively.  Using these results, we modeled the electron energy distributions of the lobes as broken power laws with age breaks. We find that a significant population of non-radiating particles is required to account for the total pressure of the eastern lobe.  In the western lobe, no such population is required and the low energy cutoff to the electron distribution there needs to be raised to obtain pressures consistent with observations. This discrepancy is a consequence of the differing X-ray photon indices, which may indicate that the turnover in the inverse-Compton spectrum of the western lobe is at lower energies than in the eastern lobe. We modeled the emission from both jets as inverse-Compton emission. There is a narrow region of parameter space for which the X-ray jet can be a relic of an earlier active phase, although lack of knowledge about the jet's electron distribution and particle content makes the modelling uncertain.

\end{abstract}



\section{Introduction}
\label{sec:intro}
Cygnus A (hereafter Cyg A) is a FRII radio galaxy \citep{Fanaroff1974}.  Its extreme radio brightness \citep{Stockton1996} made it one of the first objects of such type to be discovered. Radio observations show extended, plume-like lobe structures of synchrotron-emitting plasma, as well as jets extending to the east and west of the AGN, which terminate in bright hotspots where the jets are driving shocks into the surrounding intracluster medium (ICM) \citep{Carilli1988, Carilli1991, Carilli1994}. In the X-ray, these shocks are observed as a sharp brightness edge ahead of the hotspots, and are also referred to as the cocoon shocks \citep{Wilson2006}.

Inverse-Compton (IC) emission has been detected in the lobes and hotspots of many FRII sources,  \citep[e.g][]{ Hardcastle2002b, Konar2009}. There are different names for the IC emission depending on the seed photons that are scattered. The types usually considered are IC scattering of the Cosmic Microwave Background (IC/CMB) \citep{Harris1979}, of synchrotron photons (synchrotron self-Compton, or SSC), or of infrared photons from the AGN \citep{Brunetti1997}.  Because the IC spectrum is directly linked to the electron energy distribution, it probes low-energy electrons in the lobes. The combination of the X-ray IC spectrum, the radio synchrotron spectrum and the total pressure then provide constraints on the distribution of electron energies and the magnetic field strength in the lobes. For FRII sources it has been shown that most sources have overpressured lobes with electron-dominated internal energies \citep{Ineson2017}.  

It is common to assume equipartition to model the lobe energy density, especially for FRI galaxies. Many FRI galaxies show deficits compared to the external pressure that seem to require a significant quantity of non-radiating particles, such as protons \citep{Morganti1988, Birzan2004, Hardcastle2007}. This population could be created by entrainment of material by the jets\citep{ Croston2008, Croston2014}.

Detecting IC emission from the lobes in Cyg A has proven challenging as a result of the rich cluster environment containing bright, relatively hot thermal emission. Recent work by Wise et al. (in prep.) shows that the ICM around and in front of the western lobe is significantly hotter than on the eastern side, and shows more temperature variation. This temperature structure could be a signature of earlier cycles of AGN activity, a shock generated by the early phase merger with the northwestern subcluster (Cygnus NW), or some combination of the two.

Previous work has claimed detection of non-thermal emission in the Cyg A lobes \citep{Hardcastle2010, Yaji2010}. The non-thermal lobe fluxes in these papers are consistent with each other, although the errors are large. The result shows that Cyg A may be different from other FRII galaxies in that the electron population is not energetically dominant, and that the jet entrainment model for FRI radio galaxies may be important for Cyg A as well \citep{Hardcastle2010}. 

The wide, linear features extending east and west of the AGN in the X-ray are generally assumed to be X-ray analogues of the radio jets. Although the nature of these features is partially unclear, we will refer to them throughout this paper as the X-ray jets. \cite{Dreher1987} derived an upper limit to the thermal electron density in the lobes of Cyg A, using Faraday rotation measurements.  These limits are difficult to reconcile with a thermal model for the jet emission, as shown in \cite{SB2008}.

If the X-ray jets are non-thermal in origin, it is unclear which non-thermal emission mechanism would produce extended X-ray jets on kiloparsec scales. Generally, two different models have been invoked for these kinds of jets. The first model is the boosted IC/CMB model \citep{Tavecchio2000, Celotti2001}. In this model, high bulk Lorentz factors at small angles to the line of sight Doppler boost the upscattered photons enough to produce detectable X-ray emission. The model has been applied to quasar jets. However, in Cyg A the eastern, receding jet appears to be brighter in X-rays than the western, approaching jet. Doppler boosting would only have the opposite effect on the eastern jet, and increase the difference in intrinsic emissivities between the two jets.  Therefore, we consider the boosted IC/CMB model unlikely to apply to Cyg A. 

The synchrotron model is often proposed as an alternative to the IC/CMB model. X-ray synchrotron emission requires electrons with very high Lorentz factors. Because the lifetime of X-ray synchrotron-emitting electrons is only on the order of tens to hundreds of years, the electrons require \textit{in situ} acceleration. Electron re-acceleration is achieved through shocks, which happen locally in jet knots. A synchrotron jet model can therefore explain more naturally the knottiness seen in some of the extended X-ray jets \citep{Hardcastle2016}. In synchrotron models, many systems with multi-wavelength observations show that the radio, optical/IR, and X-ray data cannot be explained with a single electron energy distribution, which implies the existence of a second, more energetic component \citep[e.g.][]{Jester2006, Hardcastle2006, Uchiyama2006}. It is unclear how a second electron energy distribution could be created.

A few morphological differences make the Cyg A jets unlike most other radio/X-ray jet systems. The most obvious difference is that the X-ray jets are at least 4-6 times wider than the radio jet, extending several arcseconds in width. An extended X-ray jet structure has been observed  in the quasar PKS 1055+201 \citep{Schwartz2006}. Moreover, the X-ray jets and radio jets are misaligned \citep{Steenbrugge2008}. While aligned with each other close to the AGN, midway to the lobe the X-ray jets extend relatively straight towards the brightest hotspots, while the radio jets bend southwards to the fainter hotspots.   Based on these morphological differences, \cite{SB2008} argue that the Cyg A jets are IC/CMB-emitting relic jets, emitted by an older electron population that was left behind from earlier passage of the radio jet.  In this model, when the radio jet changes direction through precession or for some other reason, the electron population of the radio jet expands adiabatically into the medium, reducing electron energies to the range required to produce IC/CMB X-rays. This would explain the spatial misalignment of the jets,  as well as the greater width of the X-ray jet. The IC relic jet model could also explain the brightness difference between the two jets through the difference in light travel time. The Cyg A cocoon is inclined at $\sim 55$ degrees to our line of sight \citep{Vestergaard1993}. This means light from the eastern hotspot has an additional light travel time of $\sim 2 \times 10^{5}$ yrs. The difference in light travel time could explain the relative faintness of the western jet: it has had more time to fade and expand.

However, the question remains how this relic X-ray jet could exist long enough as a linear feature for us to observe it. If the adiabatic expansion is too fast, the jets would not be observed as a linear feature. Moreover, a fast expansion of the jet would shock the material in the lobes. The observed X-ray jet morphology implies that the jets would have to be fairly close to pressure balance with the lobes. Because the jets are brighter than the surrounding lobe, this is difficult to achieve unless the jets and lobes have significantly different electron populations or particle content. Additionally, it is difficult to maintain the observed knotty jet structure as this implies significant local pressure variations. In an expanding relic jet scenario, those pressure variations should smooth out during the expansion.

In this paper we use 1.8 Ms of new \textit{Chandra} observations, combined with 200ks of archival observations, to analyse the emission from the lobes and X-ray jets of Cyg A. Complementary results for the inner gas structure and outer lobe shocks appear in \cite{Duffy2018} and \cite{Snios2018}, respectively. We compare different models for the lobe and jet emission and constrain their parameters. With these parameters, we model the energy density of the lobes and test the possibility of an IC relic jet.  We show the data and detail the data reduction in section \ref{sec:Xraydata}. We give an overview of the statistical tools and the models that we used in section \ref{sec:statistics}. We present the results of the statistical analysis in section \ref{sec:results}, and discuss their interpretation in section \ref{sec:discussion}. We conclude in section \ref{sec:conclusion}. 

Throughout this paper, we adopt a cosmology with $H_{0}$= 69.3 km s$^{-1}$ Mpc$^{-1}$, $\Omega_{M}$ = 0.288, and $\Omega_{\Lambda} = 0.712 $ \citep{Hinshaw2013}. We use a redshift value of z=0.0561 \citep{Stockton1994}. This yields a linear scale of 66 kpc per arcminute and a luminosity distance $D_L = 253.2 $ Mpc for Cyg A. The spectral index $\alpha$ is defined so that flux $\propto \nu^{-\alpha}$, and related to the X-ray photon index as  $\Gamma = 1 + \alpha$.

\section{X-Ray Observations and Data Reduction}
\label{sec:Xraydata}

\begin{table*}
\centering
\caption{Observation log of Chandra Cyg A data used in this paper}
\begin{tabular}{p{0.8cm}p{1.4cm}p{1.0cm}p{1.7cm} | p{0.8cm}p{1.4cm}p{1.0cm}p{1.7cm}}  \hline \hline
 ObsID$^{a}$  & Date$^{b}$ & $T_{exp}$ $^{c}$ (ksec)  & Pointing $^{d}$ & ObsID$^{a}$  & Date$^{b}$ &  $T_{exp}$ $^{c}$  (ksec) & Pointing $^{d}$\\ \hline
 360* & 2000 05 21 & 34.7 & Nucleus  &17138 &2016 07 25& 26.4 & W Hotspot \\ 
 1707* & 2000 05 26 & 9.2   &Nucleus & 17513 & 2016 08 15& 49.1 & Nucleus \\
 6225 & 2005 02 15 & 24.3  &Nucleus & 17516 &2016 08 18& 49.0 & W Hotspot \\
 5831 & 2005 02 16 & 50.8 &Nucleus & 17523 &2016 08 31& 49.4& E Hotspot  \\
 6226 & 2005 02 19 & 23.7  &Nucleus & 17512 & 2016 09 15& 66.9& Nucleus  \\
 6250 & 2005 02 21 & 7.0   & Nucleus & 17139 &2016 09 16& 39.5 & W Hotspot \\
 5830 & 2005 02 22 & 23.2  &Nucleus & 17517 &2016 09 17& 26.7 & W Hotspot \\
6229 & 2005 02 23 & 22.8  & Nucleus & 19888 &2016 10 01 & 19.5 & W Hotspot \\
6228 & 2005 02 25 & 16.0   &Nucleus & 17140 &2016 10 02 & 34.3 & W Hotspot \\
6252 & 2005 09 07 & 29.7  &Nucleus  & 17507 & 2016 11 12& 32.4& Nucleus \\
17530 & 2015 04 19 & 21.3  &E Hotspot & 17520 & 2016 12 06 & 26.8 & W Hotspot \\
17650 & 2015 04 22 & 28.2  &E Hotspot  & 19956 & 2016 12 10& 54.1 & W Hotspot \\
17144 & 2015 05 03 & 49.4  &E Hotspot & 17514& 2016 12 13 &49.4 & Nucleus \\ 
17141 & 2015 08 01 & 29.6  &E Hotspot  & 17529&2016 12 15& 35.1 & E Hotspot \\
17710 & 2015 08 07 & 19.8  &E Hotspot &  17519 &2016 12 19&29.4& W Hotspot \\
17528 & 2015 08 30 & 49.3  &E Hotspot & 17135 & 2017 01 20& 19.8 & Nucleus  \\ 
17143 & 2015 09 03 & 27.1  &E Hotspot  & 17136 & 2017 01 26& 22.2 & Nucleus \\ 
17524 & 2015 09 08 & 23.0  &E Hotspot & 19996 & 2017 01 28&  28.6 & Nucleus\\
18441 & 2015 09 14 & 24.6  &E Hotspot  & 19989 & 2017 02 12& 41.5 & Nucleus\\
17526 & 2015 09 20 & 49.4  &E Hotspot & 17515 &2017 03 22& 39.0 & W Hotspot \\
17527 & 2015 10 11 & 26.8  & E Hotspot & 20043 &2017 03 26&29.3& W Hotspot \\
18682 & 2015 10 14 & 22.8  &E Hotspot  & 20044 &2017 03 27&14.6 & W Hotspot \\
18641 & 2015 10 15 & 22.4 &E Hotspot  & 17137 &2017 03 30& 25.2 & W Hotspot \\
18683 & 2015 10 18 & 15.6 &E Hotspot & 17522 &2017 04 08&48.6 & W Hotspot \\
17508 & 2015 10 28 & 14.9  &Nucleus  & 20059 &2017 04 19& 23.7 & E Hotspot \\
18688 & 2015 11 01 & 34.6 & Nucleus  & 17142 &2017 04 20& 23.3& E Hotspot \\
18871 & 2016 06 13 & 21.8   &Nucleus & 17525 &2017 04 22&24.7 & E Hotspot \\
17133 & 2016 06 18 & 30.2   &Nucleus  &  20063 &2017 04 22& 25.4 & E Hotspot \\
17510 & 2016 06 26 & 37.3  & Nucleus & 17511 &2017 05 10 & 15.9 & Nucleus \\
17509 & 2016 07 10 & 51.2  & Nucleus & 20077 &2017 05 13 & 27.7 & Nucleus \\
17518 &2016 07 16& 49.4 & W Hotspot & 20048 &2017 05 19 &22.7 & E Hotspot \\
17521 &2016 07 20& 24.5 & W Hotspot & 17134 &2017 05 20 & 29.4 & Nucleus \\ 
18886 &2016 07 23& 21.5 & W Hotspot & 20079 &2017 05 21 & 23.8 & Nucleus\\ \hline
& & & & &  Total & 2005.3 & \\  \hline \end{tabular}
\label{exposures} \\
\raggedright{
\setlength\parindent{8em}
$^{a}$The \textit{Chandra} Observation ID number. ObsID's marked with an asterisk indicate ACIS-S observations. \\
$^{b}$ The date of the observation . \\
$^{c}$ The exposure times after filtering for flares. \\
$^{d}$ The aimpoint location of the observation. Three different aimpoints have been used in this data set: the AGN, \\
as well as the western and eastern hotspots. \\}.
\end{table*}

\begin{figure*}
\centering
   \includegraphics[width=0.9\textwidth]{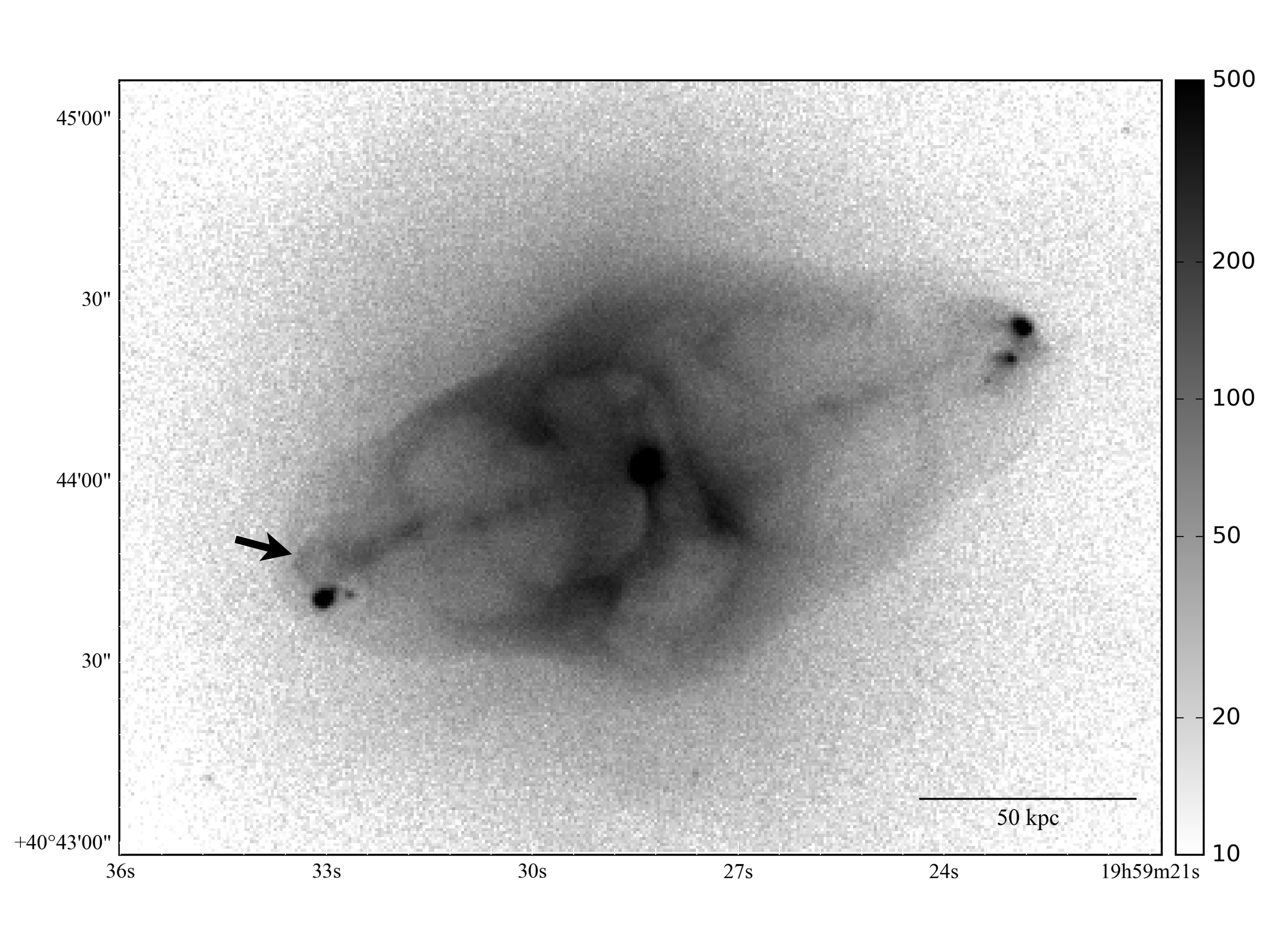}
  \caption{The merged 0.5 - 7.0 keV counts image Cyg A, binned with native 0.492 arcsec pixels. The black arrow indicates a brightness edge corresponding to the lobe edge. See text for details.}
  \label{fig:cygcore}
\end{figure*}

\subsection{Data reduction}

This paper uses nearly all of the Cyg A data available on the \textit{Chandra} archive. This includes 200 ks of previous observations taken between 2000 and 2005, and 2 Ms of recent observations taken between 2015 and 2017. A subset of 200ks of the recent observations were excluded, as they are pointed at the northwestern subcluster Cyg NW and the filamentary region between the two subclusters. This leaves a total data set of more than 2 Ms.  A log of all the observations, with their filtered exposure times and pointings, is shown in Table \ref{exposures}. For an extended review of the full data set and the large scale structure of the system, we refer to Wise et al. (in prep).

Each of these data sets has been reprocessed with \texttt{CIAO} 4.9 and \texttt{CALDB} 4.7.4 \citep{Fruscione2006}. Before reprocessing the data, we corrected for small astrometric errors caused by \textit{Chandra's} pointing accuracy of around 0.5 arcsec. We followed the procedure described by \cite{Snios2018}, briefly summarised here.  We chose ObsID 5831 as the reference observation for the high total counts. We then reprojected the event lists of the other ObsID onto the sky frame of ObsID 5831. For each ObsID, we cross-correlated a  0.5 - 7.0 keV 160 x 120 arcsec region around the central AGN with ObsID 5831 to determine the coordinate offset. The coordinate shift was then applied to the event list and aspect solution files with \textit{wcs\_update}.

After the astrometry correction, we applied the following \texttt{CIAO} processing tools. For each ObsID, a new badpix file was built with \textit{acis\_build\_badpix}. We applied the latest gain and CTI corrections with \textit{acis\_process\_events}. We created a new level 2 event file by filtering for good grades (0,2,3,4,6). After that, we filtered for GTIs with the tool \textit{deflare}. Finally, we identified readout streaks with \textit{acis\_streak\_map} and filtered them out. 

The background event files were created from the ACIS blank sky event files. The backgrounds were imported from the calibration database with the tool \textit{acis\_bkgrnd\_lookup}, and reprojected. The backgrounds were scaled to the data by using the counts between 10.0 - 12.0 keV.  The event files and backgrounds were all separately reprojected and added together to form a merged counts image and a merged background map. We show the merged 0.5 - 7.0 keV counts image in Fig. \ref{fig:cygcore}.

As well as the X-ray data, we have used two radio maps of the system: a 4.5 GHz VLA radio map from \cite{Perley1984}, and 150 MHz LOFAR radio map from \cite{McKean2016}. The radio maps were used to define the extent of the lobe extraction regions on the X-ray data. We also used the radio fluxes within these regions to do combined modelling of the radio and X-ray spectra. No additional processing has been done to the radio data. 

In the \textit{Chandra} image, we observe a brightness edge in the eastern lobe just above the northern hotspot, that corresponds with the edge of the lobe in the VLA data. This is empirical evidence that we are directly observing the non-thermal emission from the lobe in the X-ray data in this region. We have indicated this region with a black arrow in Fig. \ref{fig:cygcore}.

\subsection{Extraction regions and spectra}

\label{sec:data:spectra}
\begin{figure*}
   \includegraphics[width=1.0\textwidth]{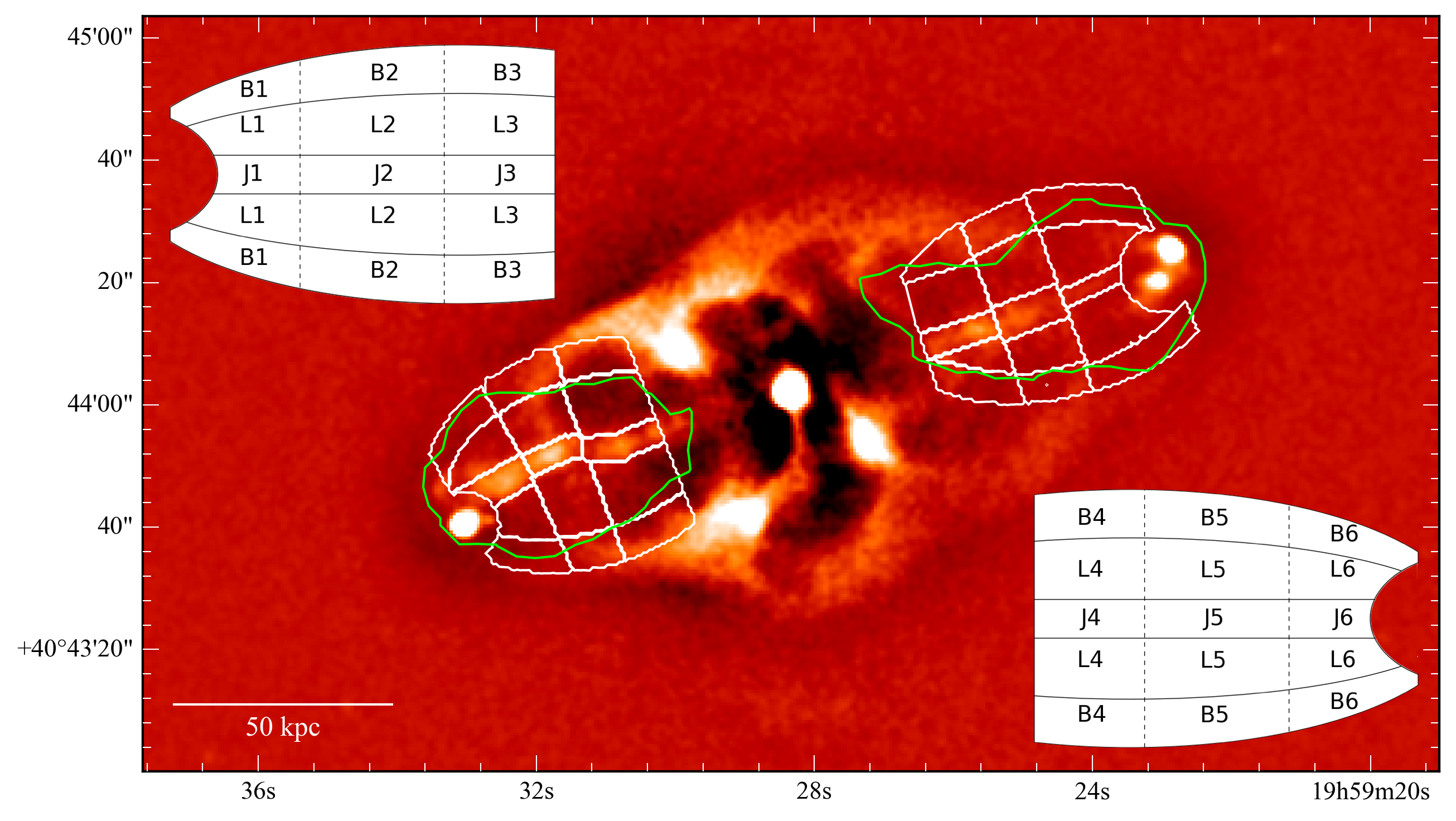}
  \caption{Radial unsharp masked residual image of the Cyg A core. Shown in green are smoothed contours of the 4.5 GHz VLA radio data. Shown in white are the jet (J), lobe (L) and background (B) extraction regions. The schematic illustrations next to each lobe indicate how the regions are labelled.}
     \label{fig:Cygregions}

\end{figure*}

\begin{figure}
   \includegraphics[width=0.5\textwidth]{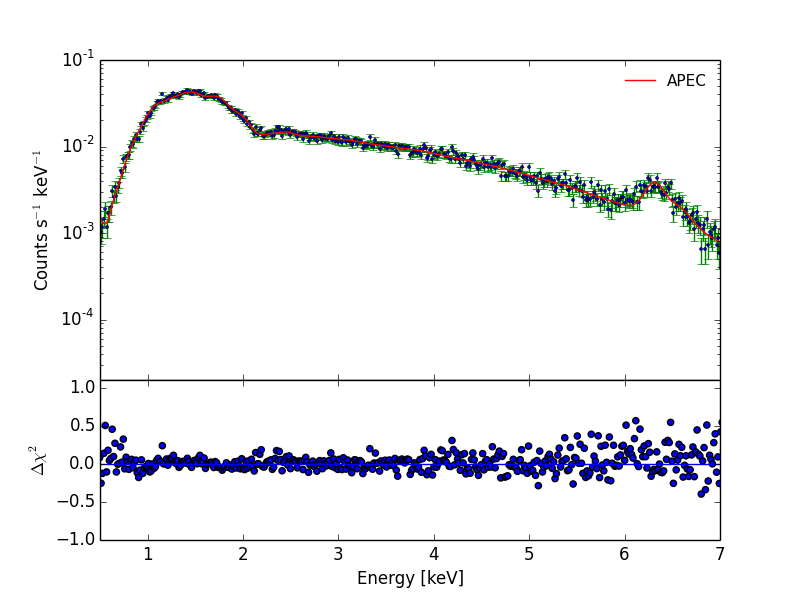}
   \includegraphics[width=0.5\textwidth]{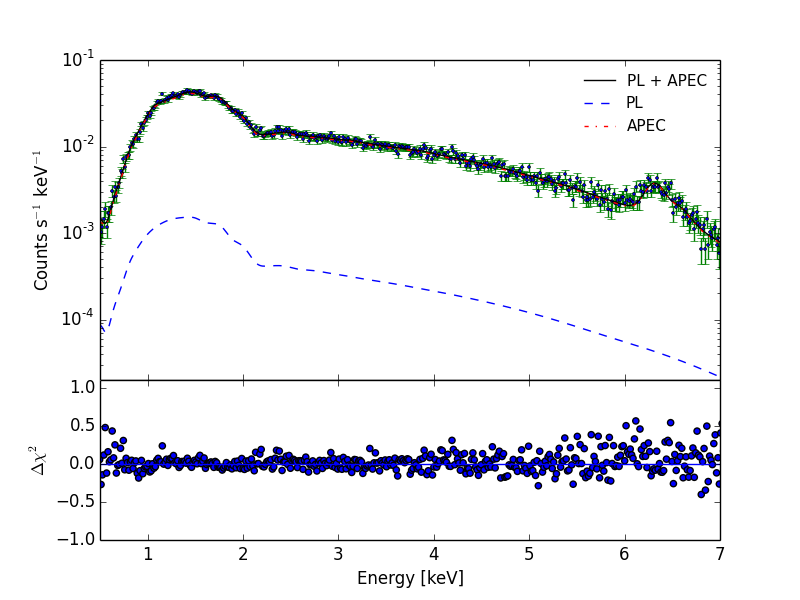}
  \caption{Comparison of two different models, both fit to the spectrum of region L2. The top image shows an \texttt{APEC} model, the bottom image a \texttt{APEC + POWERLAW} model.} 
     \label{fig:therm_nontherm}

\end{figure}

We used the \texttt{CIAO} fitting package \texttt{Sherpa} \citep{Freeman2001} to analyse the spectra. All spectral models mentioned in this paper are multiplied by a PHABS foreground Galactic absorption model. The Galactic HI column density is set at $3.1 \times 10^{21} $atoms cm$^{-2}$. This value is based on  the average of the column densities of the Leiden/Argentine/Bonn (LAB) and Dickey \& Lockman surveys \citep{Kalberla2005, Dickey1990}. The thermal model used in this paper is the Astrophysical Plasma Emission Code \citep[\texttt{APEC,}][]{Smith2001}, with the elemental abundance model from \cite{Anders1989}.

To better highlight the wealth of structure within the Cyg A cocoon shock, we created a residual map of the data. This was done by subtracting a radial unsharp masked image (Wise et al., in prep.). This technique is similar to traditional unsharp masking techniques.  A radon transform was applied to the background-subtracted, merged image of the core. Each column of pixels was then smoothed with a 7 arcsec 1D Gaussian kernel. The smoothed image was transformed back to Cartesian coordinates and subtracted from the input image. The resulting residual map, together with the extraction regions, are shown in Fig. \ref{fig:Cygregions}. The radial unsharp masking technique has the advantage that it only smooths in the radial direction. Therefore, there is less risk of creating artifacts by mismatched Gaussian smoothing kernels.

We assume that the X-ray jets are tube-like structures inside the lobes, which are in turn embedded in a shell of thermal ICM. With this geometry in mind, we define three different types of extraction regions: the X-ray jet regions (J), lobe regions (L) and thermal background regions (B). For each lobe, we defined 3 sets of J, L and B regions, which allows for variation in the thermal properties of the ICM along the jet axis. 
 
The X-ray jet regions were made to trace the jet as seen on the residual map. In the eastern lobe especially, the jet makes a noticeable bend which the extraction regions follow. The width of each jet region was set to be the FWHM of the surface brightness peak perpendicular to the jet in that region. This definition results in variations in the width of the jet extraction regions along the jet path.
 
In defining the edges of the lobe, we have taken care to include the regions with the brightest lobe emission that lie within the cocoon shock. We have therefore opted to use the VLA map rather than the LOFAR map, as electrons that produce synchrotron emission at $150$ MHz, will IC scatter those synchrotron photons to energies below $1$ eV. Because SSC emission is considered to strongly contribute to the total IC flux, we  expect regions that show only low-frequency radio emission to show less non-thermal X-ray emission.

We determined the noise level on the VLA 4.5 GHz continuum map to be 0.8 mJy/beam. We then defined contours around the lobe at the 5 $\sigma$ level, or 4 mJy/beam, at a smoothing scale of 20 pixels. These contours enclose the brightest lobe emission. The lobe regions on each side were then defined as ellipses approximately following  these contours. We trimmed the lobe regions close to the hotspots, as we expect the cocoon shock emission to dominate here over any possible non-thermal emission from the X-ray jet and lobes. In the western lobe, the ellipse was trimmed on the southern side to follow the asymmetric shape of the radio lobe. 
 
 Finally, the thermal background regions were created by drawing ellipses around the lobe regions on each side. These regions were defined close to the lobes so that their thermal properties do not differ much from the thermal properties of the material in front of the lobes.  The outer edge of the radio lobe does enter slightly into the thermal background regions in the outer parts, although the radio lobe drops off in flux sharply beyond the defined lobe size. The
thermal spectra for the background regions are not subtracted from the
spectra, but rather their temperatures and abundances are used to constrain thermal emission from material superposed on the lobe and jet regions.
 
 Using \textit{specextract}, we extracted the events within each extraction region from every ACIS chip that overlaps with that region. We obtained source spectra, response files, and blank sky spectra for each region on each ACIS chip of each observation. After the extraction, we combined the spectra of for all ACIS-S and ACIS-I observations with \textit{combine\_spectra}. This results in one combined spectrum, set of response files, and blank sky background spectrum for each region. We have used these combined spectra in the rest of the analysis. \textit{combine\_spectra} automatically adds all the exposure times of individual spectra together when combining. However, this means that when an extraction region falls on two different chips within the same ObsID, the exposure time of that ObsID is erroneously counted twice. We therefore re-calculated the exposure time of each spectrum manually after the combining process. 
 
We have chosen to combine both the ACIS-I and ACIS-S spectra together into one combined spectrum for each region. This was done because the ACIS-S data only makes up $45$ ks of the total $2$ Ms exposure time. Furthermore, the $\sim 5-8$ keV temperature of the gas around Cyg A is sufficiently high that the response below $2$ keV, where ACIS-S and ACIS-I are most different, is unlikely to drive the fit results. To test this, we created combined spectra for regions B2 and B4 that only include ACIS-I observations. We then compared the total combined spectra to the ACIS-I combined spectra. In both regions, the difference in temperature and abundance are less than $1 \%$.

In the eastern lobe and jet, the combined spectra contain an average of $\sim 70000$ and $\sim 25000$ counts per region respectively. In the western lobe and jet, the combined spectra contain an average of $\sim 36000$ and $\sim 14000$ counts per region respectively. Despite the high number of counts in each region, it is difficult to disentangle thermal from non-thermal models at CCD resolution with standard fitting procedures. To illustrate this, we took the spectrum of lobe region L2 and subtracted the blank sky background for the same region. We then fit two models to this spectrum: a thermal APEC model, or a combination of thermal and non-thermal emission (\texttt{APEC + POWERLAW}). The resulting fits are shown in Fig. \ref{fig:therm_nontherm}. An \texttt{APEC} fit gives a $\chi^2/dof$ of $433/430$, while the \texttt{APEC + POWERLAW} fit gives a $\chi^2/dof$ of $433/428$. 
 
Because the power law component is weak compared to the thermal  component, the difference in the parameters between models is small: the \texttt{APEC} model gives $ T = 6.36 \pm 0.09 $ keV and $Z = 0.42 \pm 0.03 $, while the \texttt{APEC + POWERLAW} model gives $ T = 6.48 \pm 0.18 $ keV and $Z = 0.44 \pm 0.03$, and $\Gamma = 2.06 \pm 0.62$. This example illustrates that when a non-thermal emission component is included in the model, the fit is not significantly improved, and that because of its small amplitude compared to the thermal component, it does not have a significant effect on the thermal parameters. Therefore, statistical tests such as an F-test do not give convincing evidence for or against the presence of power law emission.

Instead, of treating each region separately, we will treat all the regions in each jet and lobe together. By building a Bayesian model for each lobe and jet, we can simultaneously fit regions while setting priors for each of the parameters in our model. It also allows us to MCMC sample the models and thereby obtain posterior distributions for each parameter. We describe the statistical approach and the models used in the next section.

\section{Statistical  Analysis}
\label{sec:statistics}

\subsection{Statistical approach}
\label{sec:stats:approach}
We give a brief overview of the statistical approach here. For a more extended review of Bayesian inference, Markov Chain Monte Carlo (MCMC) sampling, and model comparison, we refer to Appendix \ref{appendix:stats}.  The models are described in more detail in section \ref{sec:Models}.

We defined two competing models for the lobe regions on each side: one model with only thermal emission and one which includes both thermal and non-thermal emission. For each of these models, we determined the maximum loglikelihood through a \textit{Maximum A Posteriori} (MAP) estimate.  Each model was also sampled with a MCMC algorithm. We used the Python module \textit{emcee} \citep{Foreman-Mackey2013}, which implements an affine invariant ensemble MCMC sampler based on \cite{Goodman2010}.  

The likelihoods obtained from the MAP estimate were used to compare the models with the corrected Akaike Information Criterion (AIC$_{\rm C}$) and the Bayesian Information Criterion (BIC).  We have used the Likelihood Ratio Test (LRT) as an additional model comparison tool. The LRT is a form of hypothesis testing for the likelihood ratio between two nested models.  The MCMC sampled data of the thermal model were used for the LRT, to generate fake data under the null hypothesis. We then applied a MAP for both models to this data, and compared the likelihood ratio to the likelihood ratio of the real data. With the help of these model comparison tests, we selected the most likely model and used the posterior distributions obtained from the MCMC sampling in the rest of the analysis. 

We defined two competing models for the X-ray jet regions on each side as well: one thermal model and one non-thermal model. Because the jets are embedded in the lobes, the jet model needs to include all the terms from the lobe model. We used the posterior distributions from the lobe models, obtained through MCMC sampling, to set priors on the lobe components in the jet regions. As in the lobes, we determined the maximum loglikelihood of the two competing models through a MAP estimate and used AIC$_{\rm C}$ and BIC to compare the models and select the most likely model. The LRT is only valid for nested models and could therefore not be used here. The most likely model was MCMC sampled and the resulting posterior distributions were used in the rest of this analysis.

Throughout this paper, when values from the posterior distributions are quoted, we have used the median together with the 14th and 86th percentile as lower and upper errors respectively. If the data are distributed as a Gaussian, this would correspond to a 1$\sigma$ credibility interval. Because the posterior distribution is not necessarily Gaussian in shape, we also show the posterior distributions that result from the MCMC sampling.

\subsection{Model description}
\label{sec:Models}

\subsubsection{Source Models}

\begin{table*}
\centering
\caption{ Description of the lobe models, $M_{\rm L0}$ and $M_{\rm L1}$, for a single lobe region.}
\begin{tabular}{ccccc}
\hline \hline
Model component & Parameter & Prior & Hyperparameters & Description  \\ \hline
Both models \\ \hline 
\\
APEC 1 & $kT_1 $ & Gaussian &   $\mu=kT_1$, $\sigma=\sigma_{kT1}$ & Plasma temperature \\
 &$Z_1 $ & Gaussian & $\mu=Z_1$, $\sigma=\sigma_{Z1}$ & Metal abundance  \\
& $Tnorm_1$ & half-Cauchy & $\mu = 0$, $\sigma = \sigma_{T1} \times A$ & Normalisation \\ 
 \\ \hdashline
 \\
   Hyperparameters &$ \sigma_{T}$\textsuperscript{a}& uniform &  $10^{-7}  <  \sigma_{T} < 10^{-3} $ & $\sigma$ of  $Tnorm_1$ half-Cauchy \\
 \\ \hline
   $M_{\rm L1}$ \\ \hline
   \\
POWERLAW 1 & $\Gamma_{1}$   \textsuperscript{a} & uniform &  $1.0 < \Gamma$  & Photon index \\ 
&$Pnorm_1 $    & half-Cauchy &   $\mu = 0$, $\sigma = \sigma_{L} \times A$ & Normalisation  \\ 
    \\ \hdashline
    \\
Hyperparameters &$ \sigma_{L}$\textsuperscript{a} & uniform & $10^{-7}  <  \sigma_{L} < 10^{-3} $                                  \textsuperscript{b} & $\sigma$ of $Pnorm_1$ half-Cauchy  \\
 \\  \hline
\label{tab:lobemodels}

\end{tabular}

\raggedright{\setlength\parindent{14em} \textsuperscript{a} These parameters are linked between regions in the same lobe. See text for details. }\\
\end{table*}

\begin{table*}
\centering
\caption{Description of models $M_{\rm J0}$ and $M_{\rm J1}$, for a single jet region.}
\begin{tabular}{ccccc}
\hline \hline
Model component & Parameter & Prior & Hyperparameters  & Description\\ \hline
Both models \\ \hline 
   \\
   $M_{\rm L}$ \textsuperscript{a} & & & & Model $M_{\rm L0}$ or $M_{\rm L1}$ \\
 \\  \hline
   $M_{\rm J0}$ \\ \hline 
       \\
APEC 2 & $kT_2 $ \textsuperscript{b} & uniform &   $1.0 < kT_2 < 10.0$ & Plasma temperature  \\
 &$Z_2$ \textsuperscript{b} & uniform & $ 0.0 < Z_2 < 1.0 $  & Metal abundance  \\
& $Tnorm_2$ & half-Cauchy & $\mu = 0$, $\sigma = \sigma_{T2} \times A$  & Normalisation \\ 
\\ \hdashline 
\\
Hyperparameters &$ \sigma_{J}$\textsuperscript{a} & uniform &  $10^{-7} <  \sigma_{J} < 10^{-3}  $   \textsuperscript{b}  & $\sigma$ of $Tnorm_2$ half-Cauchy \\ 
\\  \hline
 $M_{\rm J1}$  \\ \hline 
\\
POWERLAW 2 &  $\Gamma_2 $   \textsuperscript{b} & uniform & $1.0 < \Gamma$  & Photon index \\ 
&$ Pnorm_2 $    & half-Cauchy & $\mu = 0$, $\sigma = \sigma_{J} \times A $  & Normalisation \\ 
\\ \hdashline
\\
Hyperparameters &$ \sigma_{J}$\textsuperscript{b} & uniform & $10^{-7}  <  \sigma_{L} < 10^{-3} $                                  \textsuperscript{b} & $\sigma$ of $Pnorm_2$ half-Cauchy \\ 

 \\  \hline
\label{tab:jetmodels}

\end{tabular}

\raggedright{\setlength\parindent{14em} \textsuperscript{a} The priors for each parameter are obtained from the posterior distributions of the lobe models.\\ The normalisation priors are scaled by the lobe/jet area ratio. See text for details.  \\
\textsuperscript{b} These parameters are linked between regions in the same lobe. See text for details.}
\end{table*}

The lobes and X-ray jets were analysed sequentially, so that we can apply model comparison tests first to the lobes and then to the jets. The background regions are not included in the model itself. Instead, they are fit in \textit{Sherpa} with an \texttt{APEC} model, and the temperatures and abundances from these regions are used as priors for the thermal components in the lobe and jet models, as described in more detail in section \ref{sec:stats:priors}. Because the definition of the edge of the lobe is somewhat arbitrary, we cannot rule out that some non-thermal emission is present in the background regions as well. However, as we have already seen in the fit comparison in section \ref{sec:data:spectra}, adding a non-thermal component to the model affects the temperature and abundance very little, even in a region that is in the middle of the lobe. Therefore, the error in our assumption will likely be smaller than the width of the prior. 

Each spectrum is fitted between 0.5 and 7.0 keV. In the lobe regions, we compare two different models. In model $M_{\rm L0}$, every region contains a thermal model. The alternative model, $M_{\rm L1}$ contains the same thermal model, with a power law added to describe the non-thermal emission. The models are nested, such that $M_{\rm L1}$ = $M_{\rm L0}$ when the amplitude of the power law is zero. Lobe models $M_{\rm L0}$ and $M_{\rm L1}$ are described in Table \ref{tab:lobemodels}. We give a detailed description on the priors of the model in section \ref{sec:stats:priors}.

Because we assume that the jets are embedded in the lobes, the jet model consists of lobe model $M_{\rm L}$ with an additional component to describe the jet. Whether the background is the thermal model $M_{\rm L0}$ or the model with non-thermal emission $M_{\rm L1}$is determined in the model selection between $M_{\rm L0}$ and $M_{\rm L1}$. The jet itself is modeled either as a  second thermal component with the same abundance and temperature across the jet, $M_{\rm J0}$, or a second  power law with the same photon index across the jet, $M_{\rm J1}$. The jet models $M_{\rm J0}$ and $M_{\rm J1}$ are described in Table \ref{tab:jetmodels}. The jet models, in contrast to the lobe models, are not nested. This means that we can apply the information criteria but not the likelihood ratio test. 

We note that Tables \ref{tab:lobemodels} and \ref{tab:jetmodels} show the parameters for just one set of lobe and jet regions. As indicated in the table, most parameters are different for each lobe and jet region. However, we have constrained the photon index $\Gamma_1$ and the parameters $\sigma_L$ and $\sigma_J$ only have one value throughout the lobe. In the jet models, we assume either a single temperature and abundance ($M_{\rm J0}$), or a single photon index $\Gamma_2$ ($M_{\rm J1}$) throughout the entire jet. 

\subsubsection{Priors}
\label{sec:stats:priors}

From MCMC sampling of the lobe models we obtain posterior distributions of each parameter in the model. We subsequently use these posterior distributions as priors for the jet model.  The probability distributions are obtained by making a 1-dimensional Kernel Density Estimation (KDE) over the posterior distribution of each parameter. We have used an asymmetric KDE so that the smoothing effect close to the prior boundaries is minimized. This is a particular concern for some of the normalisation parameters, where most of the posterior distribution could lie close to the prior boundary at zero.

We make the assumption that any thermal or non-thermal model component in a given lobe region has the same surface brightness as that same component in the corresponding jet region. This allows us to take the posterior distributions of the thermal or non-thermal normalisation in a lobe region, scale them by area and use them as prior distributions for the jet region. Because the jet models either contain two different thermal or non-thermal model components, setting a prior on one of them makes it easier to distinguish between these two components. The assumption that the surface brightness of a component is the same in the middle of the lobe (where the X-ray jet is) as on the side, is unlikely to be completely accurate. However, assuming anything about the 3-dimensional geometry of the lobe and the jet would introduce additional uncertainties as well. Furthermore, the error in this assumption will likely be subsumed in the width of the input prior distribution. 

The models use a few prior distributions which require additional parameters. Parameters of a prior distribution are referred to as \textit{hyperparameters}.  Hyperparameters can be fixed, as is the case for e.g. the peak and width of the Gaussian prior distribution on the temperature and abundance. In the case of the half-Cauchy distributions on the normalisations however, we include the hyperparameter as a free parameter in the model. The hyperprior used for those hyperparameters is a uniform prior with upper and lower bounds.

For the temperature and abundances of the foreground ICM, we have assumed a Gaussian prior with a mean $\mu$ and a standard deviation $\sigma$. These parameters are set by performing an \texttt{APEC} fit to the background region adjacent to the lobe region. The results from these fit are then used to set $\mu$ and $\sigma$ of the corresponding temperature and abundance prior distribution.

The normalisation parameters have a half-Cauchy prior distribution. This distribution includes the zero point, is heavy-tailed, and is  recommended for scale parameters in Bayesian hierarchical models \citep{Gelman2006, Polson2011}. It is especially important that the normalisation prior includes zero, to satisfy the requirement of nested models.

While the peak of the half-Cauchy distribution is at zero, the $\sigma$ of the distribution is a hyperparameter that is sampled in the model.  We have opted to use one $\sigma$ for each model component throughout all the regions, scaled by the area of that region. This was done to make each model component comparable between regions. To take $\sigma_L$ as an example: if there is any non-thermal emission in the lobe regions, we would not expect there to be a lot of non-thermal emission in L1, none in L2, and a lot in L3. We therefore use one value of $\sigma_L$, scaled by area, that sets the half-Cauchy prior distributions on the non-thermal normalisations in all three regions. The three different model components each have one hyperparameter: $\sigma_T$, $\sigma_L$, $\sigma_T$. The area is indicated in Tables \ref{tab:lobemodels} and \ref{tab:jetmodels} as $ A$.  

Finally, we set a uniform prior on the photon index $\Gamma$, with a lower bound of 1. This is because we require the power law to go down with increasing energy. We note again that $\Gamma$ is held constant throughout both the lobe and the X-ray jet.

We can now write out the full posterior equations for lobe models $M_{\rm L0}$ and $M_{\rm L1}$ and jet models $M_{\rm J0}$ and $M_{\rm J1}$, for lobe spectra $D_l$ and jet spectra $D_j$:
\begin{equation}
\label{fullpost_L0}
\begin{aligned}
p(\Theta   \, \big{|}  \, \{D_{l}\}_{l=1}^{L}  , M_{\rm L0}) \propto \\ 
\bigg{(} \prod_{l=1}^{L} p(D_l \, \big{|} \, kT_1, Z_1, Tnorm_1) \; p(kT_1 \, \big{|} \, \mu_{kT}, \sigma_{kT})  \\ 
\,  \; p(Z_1 \, \big{|} \, \mu_{Z}, \sigma_{Z} ) \;  p(Tnorm_1 \, \big{|} \, \sigma_T \times A)  \bigg{)} p(\sigma_T \, \big{|} \, min, max) , 
\end{aligned}
\end{equation}
\begin{equation}
\label{fullpost_ML1}
\begin{aligned}
p(\Theta   \, \big{|}  \, \{D_{l}\}_{l=1}^{L} ,M_{\rm L1})  \propto \\ \bigg{(} \prod_{l=1}^{L} p(D_l \, \big{|} \, kT_1, Z_1, Tnorm_1, \Gamma_1, Pnorm_{1}) \; p(kT_1 \, \big{|} \, \mu_{kT}, \sigma_{kT}) \\ 
 \; p(Z_1 \, \big{|} \, \mu_{Z}, \sigma_{Z} ) \;  p(Tnorm_1 \, \big{|} \, \sigma_T \times A)  \;  p(Pnorm_{1} \, \big{|} \,  \sigma_L \times A) \bigg{)} \\
 \bigg{(} p( \Gamma_1 \, \big{|} \, min, max) \; p(\sigma_T \, \big{|} \, min, max) \bigg{)} ,
\end{aligned}
\end{equation}
\begin{equation}
\label{fullpost_J0}
\begin{aligned}
p(\Theta   \, \big{|}  \,  \{D_{j}\}_{j=1}^{J},M_{\rm J0})  \propto \\ 
\bigg{(} \prod_{j=1}^{J} p( D_j \, \big{|} \,  M_{\rm L}, kT_2, Z_2, Tnorm_{2}) 
\; p(M_{\rm L})  \; p(Tnorm_{2} \, \big{|} \,   \sigma_{T2} \times A) \bigg{)} \\ 
\; \bigg{(} p(kT_2 \, \big{|} \, min, max) \;  p(Z_2 \, \big{|} \, min, max) \; p(\sigma_{T2} \, \big{|} \, min, max)\bigg{)} 
\end{aligned}.
\end{equation}
\begin{equation}
\label{fullpost_J1}
\begin{aligned}
p(\Theta   \, \big{|}  \,  \{D_{j}\}_{j=1}^{J},M_{\rm J1})  \propto \\ 
\bigg{(} \prod_{j=1}^{J} p( D_j \, \big{|} \,  M_{\rm L}, \Gamma_2, Pnorm_{2}) \; 
p(M_{\rm L})  \; p(Pnorm_{2} \, \big{|} \,   \sigma_J \times A) \bigg{)} \\ 
\; \bigg{(} p(\Gamma_2 \, \big{|} \, min, max) \;  p(\sigma_J \, \big{|} \, min, max) \bigg{)} ,
\end{aligned}
\end{equation}

The above equations are Bayes' theorem with all likelihood and prior terms written out, for both lobe and both jet models. The first term after each product sign is the Poisson likelihood for that region. The other terms describe the prior for each of the parameters.  The last terms in each equation are the priors for parameters that are linked between the regions: the photon indices $\Gamma_1$ and $\Gamma_2$, the temperature and abundance in the thermal jet model ($M_{\rm J1}$), and the hyperparameters $\sigma_T$, $\sigma_L$ and $\sigma_J$.

Each lobe contains 3 lobe and 3 jet regions. We analysed the eastern and western lobes individually, so per lobe we have 10 free parameters for $M_{\rm L0}$, 14 free parameters for $M_{\rm L1}$, 19 free parameters for $M_{\rm J0}$ and 18 for $M_{\rm J1}$. 

\subsubsection{Blank sky backgrounds}

Because we are using Poisson likelihoods, we cannot subtract the ACIS blank sky background spectra from the source spectra. This would make the resulting data non-Poissonian. 

The ACIS blank sky backgrounds are made by averaging the backgrounds of many different ObsIDs, for each pixel on each ACIS chip. The backgrounds include both instrumental and sky components. The instrumental background can be described by a continuum and multiple fluorescent emission lines \citep{Bartalucci2014}. The sky background consists mainly of diffuse Galactic thermal emission and a weakly absorbed power law \citep{Hickox2006}. Therefore, any physical model that describes the full background will have to consist of many different components. Furthermore, because of the large number of ObsIDs involved in making the backgrounds, it is not possible to create separate spectral response files for the blank sky backgrounds with the usual CIAO tools.  

In our regions of interest, the contribution of the blank sky backgrounds are rather small, making up between 0.3 and 0.9\% of the total counts. We have therefore opted to model the blank sky spectra for each region parametrically rather than with a physical model. For each background spectrum, we group the bins to have a minimum of 25 counts in each bin, depending on the number of counts in the background spectrum. After grouping, we interpolate a non-smoothed quadratic spline through the grouped spectrum with \textit{scipy.interpolate.UnivariateSpline}. The spline is then scaled with the exposure time ratio between the source and background spectra. 

We then take this spline to be the background model $m_{BG}$  for that region, and we add it to the model before calculating the Poisson loglikelihood as
\begin{equation}
m_{\rm full}(\Theta) = m_{\rm data}(\Theta) + m_{\rm BG}.
\end{equation}

\section{Results}
\label{sec:results}

\subsection{Model comparison}

We first apply the AIC$_{\rm C}$, BIC and Likelihood ratio test to the lobe regions, and select between model $M_{\rm L0}$ and $M_{\rm L1}$. We then move on to the jet regions, where we use either $M_{\rm L0}$ and $M_{\rm L1}$ as part of the jet model and select between $M_{\rm J0}$ and $M_{\rm J1}$. 

\subsubsection{The lobes}

\begin{table}
\centering
\caption{AIC$_{\rm C}$ and BIC model comparison for the lobe regions.}
\begin{tabular}{ccccccc}
\hline \hline
Lobe & $k_{M_{\rm L0}}$\textsuperscript{a} & $k_{M_{\rm L1}}$\textsuperscript{b}& n\textsuperscript{c} & $ ln(\frac{\hat{L_1}}{\hat{L_0}})$\textsuperscript{d}  & $\Delta AIC_C $  \textsuperscript{e} & $\Delta BIC $ \textsuperscript{f} \\ \hline
\textbf{East} & 10 & 14  & 1437 & 23.8 &  -39.6 & -18.8  \\
\textbf{West} & 10 & 14 & 1437 & 18.9 & -29.8 & -9.0 \\
\hline
\label{tab:lobeIC} 
\end{tabular}
\raggedright{\setlength\parindent{0em} \textsuperscript{a} The number free parameters in $M_{\rm L0}$. \\
\textsuperscript{b} The number of free parameters in $M_{\rm L1}$. \\
\textsuperscript{c} The total number of data points (i.e. spectral channels). \\
\textsuperscript{d} The log-likelihood ratio of the best-fit likelihoods to $M_{\rm L0}$ and $M_{\rm L1}$. \\
\textsuperscript{e} The difference in the AIC$_{\rm C}$ between $M_{\rm L0}$ and $M_{\rm L1}$. Positive values indicate evidence in favour of $M_{\rm L0}$, negative values indicate evidence in favour of $M_{\rm L1}$. \\
\textsuperscript{f} As above, for the BIC. 
}
\end{table}

\begin{figure}
   \includegraphics[width=0.45\textwidth]{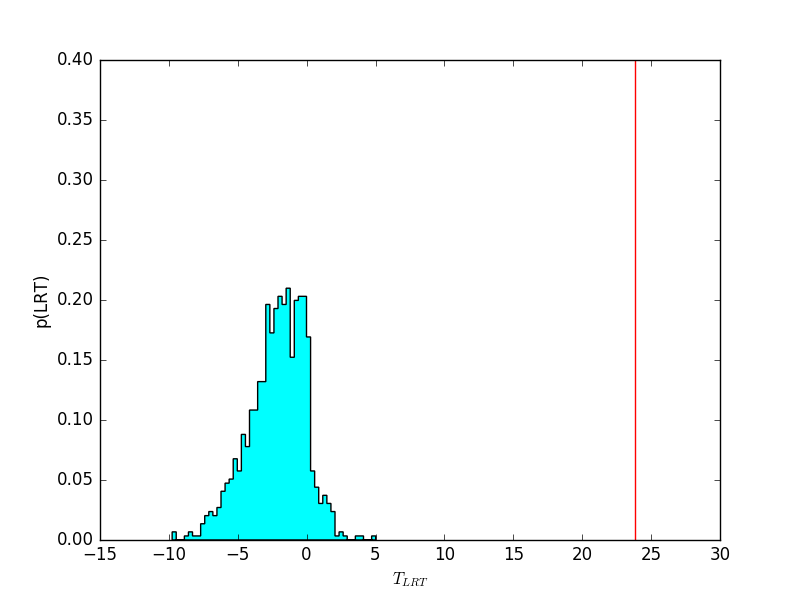}
    \includegraphics[width=0.45\textwidth]{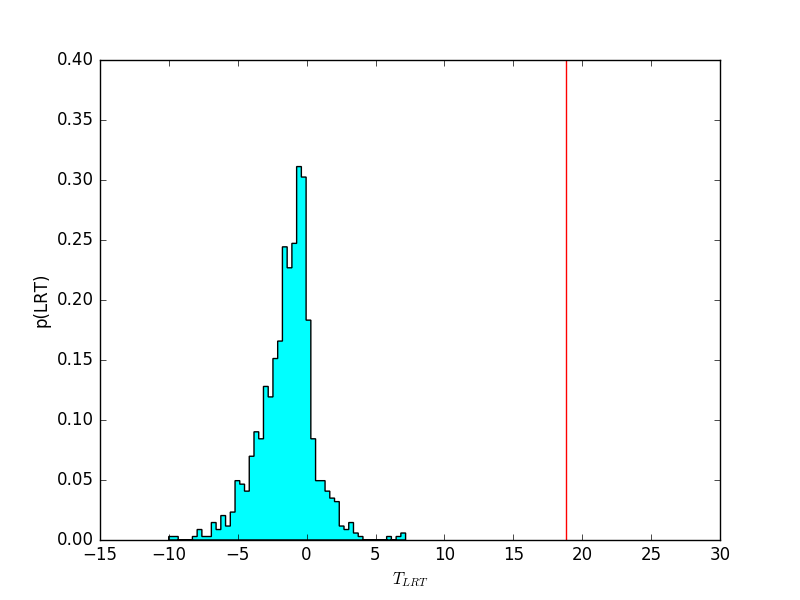}
  \caption{Top: Likelihood-ratio distribution for $M_{\rm L0}$ and $M_{\rm L1}$ in the eastern lobe. The likelihood ratio of the real data is shown as a red line.
Bottom: As above, for the western lobe.} 
    \label{fig:lobe_LRT}
\end{figure}

In the lobes, we compared a model with a single thermal component, $M_{\rm L0}$, with a model containing both a thermal and non-thermal component ,$M_{\rm L1}$. The AIC$_{\rm C}$ and BIC values are shown in Table \ref{tab:lobeIC}. We refer to Table \ref{tab:Jeffreys} for the relation between AIC$_{\rm C}$ or BIC and the strength of the evidence for or against a given model. Both information criteria show a strong preference for model $M_{\rm L1}$. It should be noted that the number of data points $n$ may be overestimated, as the number of spectral bins is higher than the spectral resolution. Thus, adjacent spectral bins are strongly correlated and the number of spectral bins is not an accurate representation of the sample size n. To get around this problem, the concept of an 'effective sample size' is used in some fields \citep[e.g.][]{Thiebaux1984}. However, lowering the sample size will skew the results even more towards model $M_{\rm L1}$, and it will therefore not influence our conclusion of $M_{\rm L1}$ as the most likely model.

The results of the likelihood ratio test are shown in Fig. \ref{fig:lobe_LRT}.  For each lobe, we calculated the likelihood ratio 1000 times. We find that none of the simulated likelihood ratios come close to the real likelihood ratio, which means that the p-value in both lobes is significantly smaller than 0.001. Together with the information criteria, we have strong evidence in favour of model $M_{\rm L1}$. This confirms the presence of non-thermal X-ray emission in the lobes at high significance. $M_{\rm L1}$ will be used as part of the jet model in all subsequent analysis.

\subsubsection{The X-ray jets}

\begin{table}
\centering
\caption{AIC$_{\rm C}$ and BIC model comparison for the jet regions.}
\begin{tabular}{ccccccc}
\hline \hline
Lobe & $k_{M_0}$\textsuperscript{a} & $k_{M_1}$ \textsuperscript{b}& n \textsuperscript{c} & $ ln(\frac{\hat{L_1}}{\hat{L_0}})$  \textsuperscript{d}& $\Delta AIC_C $ \textsuperscript{e}& $\Delta BIC $  \textsuperscript{f}\\ \hline
\textbf{East} & 19 & 18  & 1437 & -1.0 &  -0.2 & -5.5  \\
\textbf{West} & 19 & 18 & 1437 & 1.5 & -5.0 & -10.3\\
\hline

\label{tab:jetIC} 

\end{tabular}
\raggedright{\setlength\parindent{0em} \textsuperscript{a} The number free parameters in $M_{\rm J0}$. \\
\textsuperscript{b} The number of free parameters in $M_{\rm J1}$. \\
\textsuperscript{c} The total number of data points (i.e. spectral channels). \\
\textsuperscript{d} The log-likelihood ratio of the best-fit likelihoods to $M_{\rm J0}$ and $M_{\rm J1}$. \\
\textsuperscript{e} The difference in the AIC$_{\rm C}$ between $M_{\rm J0}$ and $M_{\rm J1}$. Positive values indicate evidence in favour of $M_{\rm J0}$, negative values indicate evidence in favour of $M_{\rm J1}$. \\
\textsuperscript{f} As above, for the BIC.
}
\end{table}

\begin{figure}
   \includegraphics[width=0.5\textwidth]{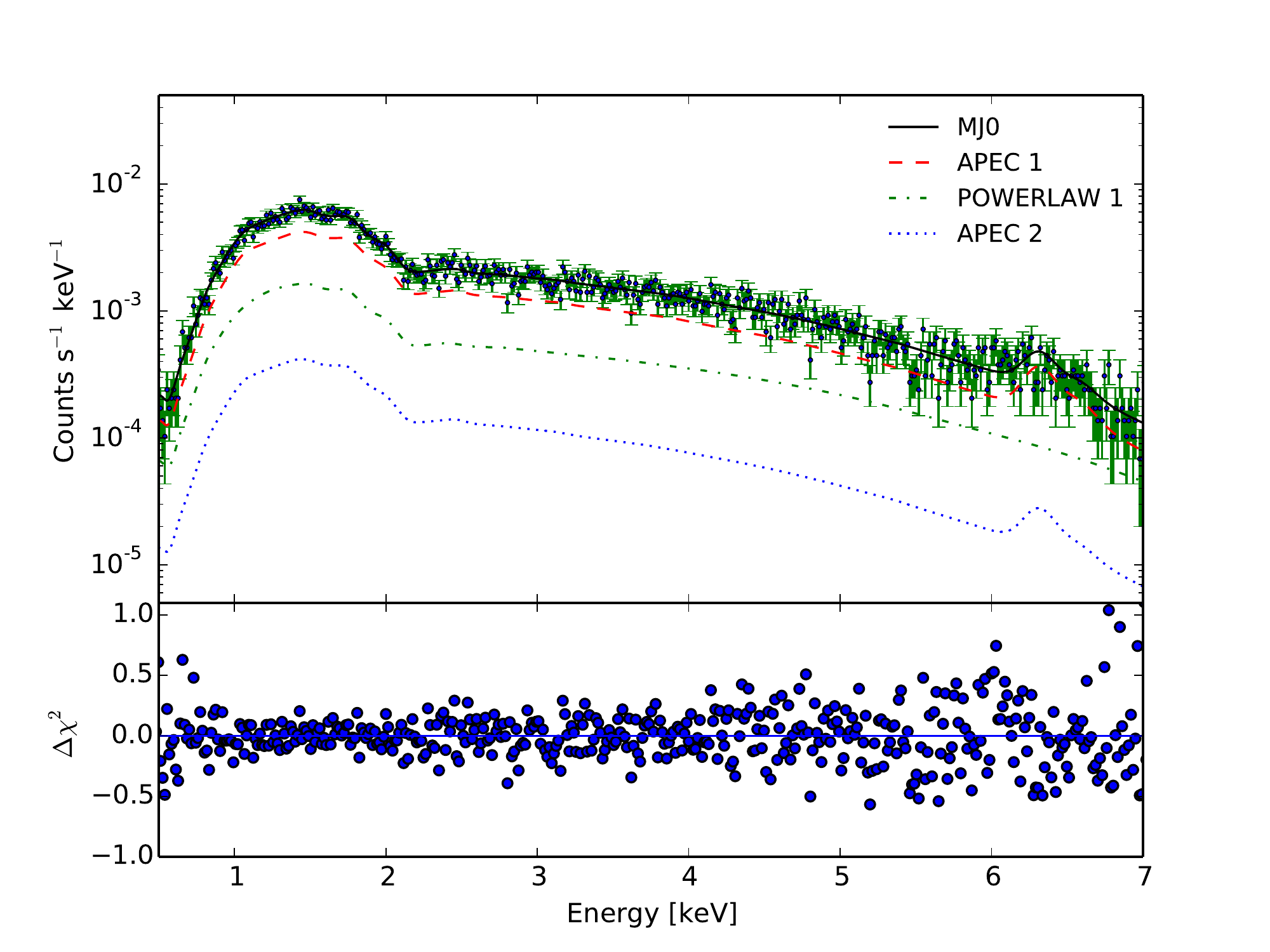}
   \includegraphics[width=0.5\textwidth]{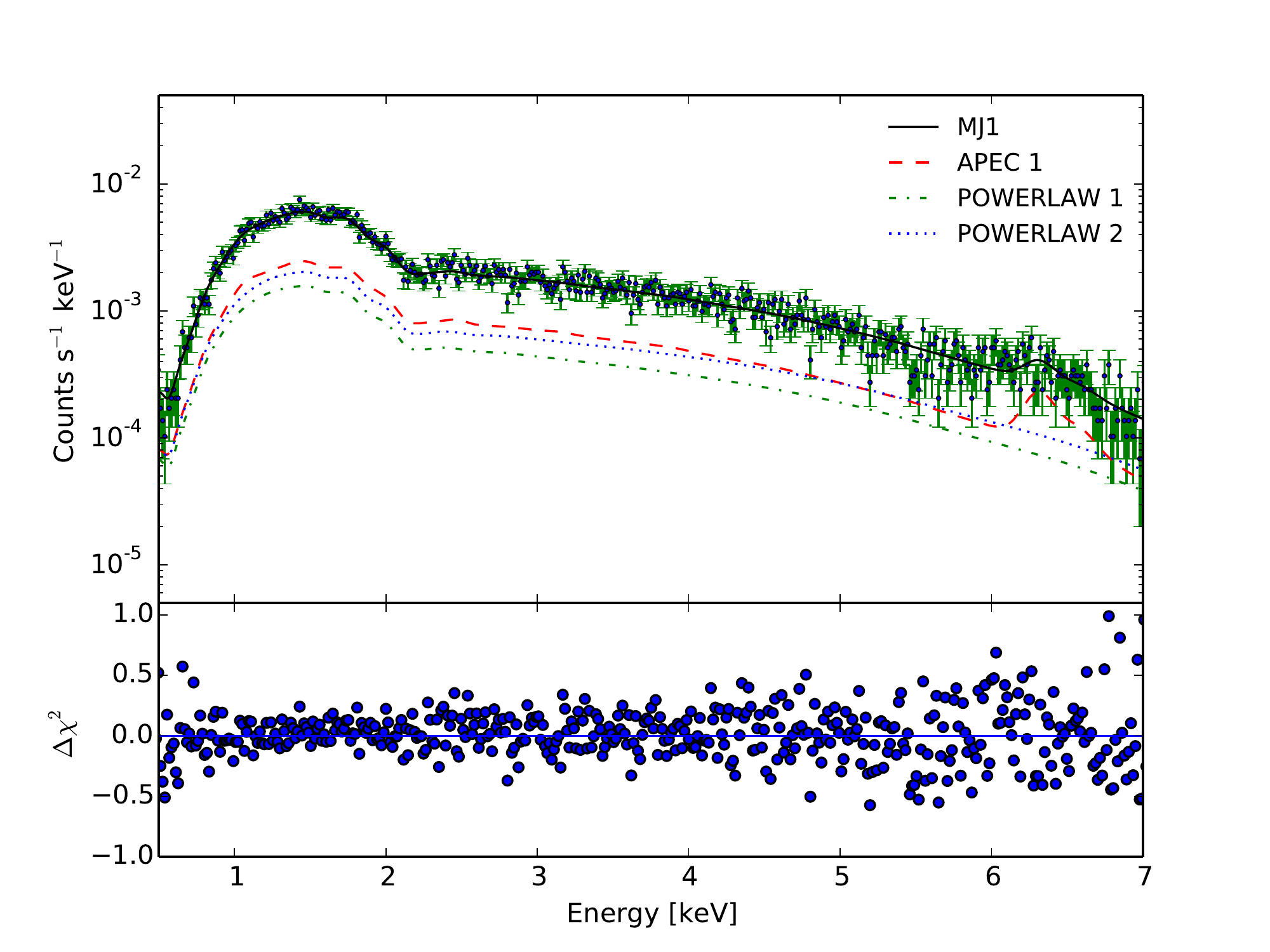}
  \caption{Comparison of models $M_{\rm J0}$ (top) and $M_{\rm J1}$ (bottom) in region J1.} 
     \label{fig:MJ0_MJ1}

\end{figure}

The model comparison tests in the lobe regions strongly prefer model  $M_{\rm L1}$ over $M_{\rm L0}$. Therefore, in the jet regions, we compared two different models with 3 components, of which the first two components are the thermal and non-thermal components of $M_{\rm L1}$. We added either a thermal component, $M_{\rm J0}$, or a non-thermal component, $M_{\rm J1}$, to model the jet emission. The results of the model comparison between $M_{\rm J0}$ and $M_{\rm J1}$, with AIC$_{\rm C}$ and BIC, are listed in Table \ref{tab:jetIC}. 

In the eastern jet, the likelihood ratio $ln(\frac{\hat{L_1}}{\hat{L_0}})$ is negative, meaning that the thermal model $M_{J0}$ has a higher likelihood than the non-thermal model $M_{J1}$. However, because $M_{J0}$ has more parameters, BIC prefers model $M_{J1}$ and AIC$_{\rm C}$ is close enough to zero as to be inconclusive. In the western jet,  both information criteria show clear preference for model $M_{\rm J1}$. We note that, as in the model selection of the lobes, the number of spectral channels $n$ might not be a correct representation of the amount of independent data points because the width of each channel is smaller than the spectral resolution of ACIS. Lowering $n$ would bring the AIC$_{\rm C}$ and BIC values closer to each other, skewing the AIC$_{\rm C}$ further towards $M_{\rm J1}$ and the BIC towards $M_{\rm J0}$. For example, setting $n=300$, will yield $\Delta$AIC $=-1.2$ and $\Delta$BIC $=-3.9$ in the eastern jet and $\Delta$AIC $=-6.0$ and $\Delta$BIC $=-8.7$ in the western jet, indicating moderate to strong evidence for $M_{\rm J1}$. 

In the eastern jet, at the maximum posterior of model $M_{\rm J0}$, we find that a thermal X-ray jet would have $T = 5.7$ keV and $Z = 0.21$ Z$_\odot$. However, the spectral normalisation of the jet component is lower than what would be expected based on the surface brightness.  For example, the spectral normalisation of the jet in region J1 is about 10 times lower than that of the ICM thermal component.  In the western jet, a thermal jet would have $T = 8.0$ keV and $Z  = 0.10$ Z$_\odot$, but the spectral normalisation is multiple orders of magnitude below the ICM thermal component,  or effectively zero.  

Fig. \ref{fig:MJ0_MJ1} shows a comparison of models $M_{\rm J0}$ and $M_{\rm J1}$ in jet region J1, illustrating that the jet component has much  lower normalisation in  $M_{\rm J0}$.  In both the eastern and the western side, the emission from the jet is mostly attributed to the ICM thermal emission.  Optimizing the posterior does not produce reasonable parameters for a thermal jet model.  Combined with the results from the AIC$_{\rm C}$ and BIC, which all point in favour of model $M_{\rm J1}$, we conclude that the favoured model includes a non-thermal emission component to describe the jet emission.  In the analysis that follows, we will therefore use model $M_{\rm L1}$ in the lobe regions and $M_{\rm J1}$ in the jet regions.

\subsection{Thermal emission components}
\label{sec:results:thermal}
\begin{table}
\centering
\caption{The thermal properties of the background, lobe and jet regions. Background regions were fit with a \texttt{PHABS x APEC} model. The temperatures and abundances are shown with $1 \sigma$ errors. The temperatures and abundance posterior distributions of the lobe and jet regions are taken from models $M_{\rm L1}$ and $M_{\rm J1}$. }
\begin{tabular}{cccc}
\hline \hline
Region & kT (keV)  & $ Z \, (Z_\odot)$ & $\chi^2$/dof \\ \hline 
\textbf{B1}& $6.80 \pm 0.30$  & $0.41 \pm 0.05 $ &  $524.12 \, / \,  521$ \\
\textbf{L1} & $6.54_{-0.16}^{+0.16}$ & $0.47_{-0.05}^{+0.05}$ &   \\ 
\textbf{J1} & $6.53_{-0.17}^{+0.17}$ &  $0.46_{-0.05}^{+0.05}$ & \\ \hdashline
\rule{0pt}{3ex}  \textbf{B2} & $6.30 \pm 0.11 $ & $0.61 \pm 0.04$ & $684.04\, / \, 625$  \\
\textbf{L2} & $6.35_{-0.10}^{+0.10}$ & $0.55_{-0.04}^{+0.04}$ & \\
\textbf{J2} & $6.36_{-0.11}^{+0.11}$ & $0.55_{-0.04}^{+0.04}$ & \\ \hdashline
\rule{0pt}{3ex} \textbf{B3}& $5.48 \pm 0.12 $ & $0.68 \pm 0.04 $ & $867.23 \, / \, 812 $ \\
\textbf{L3} & $5.77_{-0.08}^{+0.08}$ & $0.72_{-0.03}^{+0.03}$ & \\
\textbf{J3} & $5.79_{-0.09}^{+0.09}$ & $0.71_{-0.03}^{+0.03}$ & \\ \hdashline
\rule{0pt}{3ex} \textbf{B4} & $5.91 \pm 0.13 $ & $ 0.52 \pm 0.04 $ & $805.63 \, / \, 774$  \\
\textbf{L4} & $5.97_{-0.20}^{+0.20}$ & $0.60_{-0.06}^{+0.06}$ & \\
\textbf{J4} & $6.06_{-0.21}^{+0.20}$ & $0.58_{-0.05}^{+0.05}$& \\ \hdashline
\rule{0pt}{3ex} \textbf{B5} & $6.80 \pm 0.20 $ & $0.49 \pm 0.05 $& $720.10 \, / \, 696 $ \\
\textbf{L5} &$7.19_{-0.32}^{+0.38}$ & $0.53_{-0.06}^{+0.07}$ & \\
\textbf{J5} &$7.02_{-0.25}^{+0.34}$ & $0.59_{-0.06}^{+0.07}$ & \\ \hdashline
\rule{0pt}{3ex} \textbf{B6} & $7.72 \pm 0.26 $ & $0.38 \pm 0.05 $ & $707.86 \, / \, 711$ \\
\textbf{L6} & $8.03_{-0.43}^{+0.46}$&  $0.41_{-0.07}^{+0.08}$& \\
\textbf{J6} &$8.00_{-0.43}^{+0.46}$ & $0.42_{-0.07}^{+0.08}$ & \\ 
\hline

\label{tab:therm} 

\end{tabular}
\end{table}

The results for the thermal properties of the background, lobe and jet regions are listed in Table \ref{tab:therm}.  As described in section \ref{sec:Models},  the temperatures and abundances of the background regions were obtained by fitting a  \texttt{PHABS X APEC} model to the spectra with \textit{Sherpa}. The obtained values were then used as priors for the lobe model, and the distributions from the lobe models were in turn used as priors for the jet model. Table \ref{tab:therm} shows that none of the temperatures and abundances within a set of background, lobe and jet regions deviate significantly from each other. 

Consistent with \cite{Snios2018} and Wise et al. (in prep), we observe that the temperatures increase with distance from the AGN and significantly higher temperatures on the western side than on
the east Our results are also broadly consistent with the temperatures found by\cite{Wilson2006} in the regions around the lobe.

The inner background regions B3 and B4 are just on the edge of the bright, rib-like structures extending outward from the AGN. \cite{Duffy2018} suggest that these rib-like structures are a result of the destruction of the cool core during initial passage of the radio jet. Thermal plasma from the cool core would then be pushed into a cylindrical rib-like shape by backflow antiparallel to the direction of the jet.  The ribs have temperatures of around 2.5 - 4.5 keV \citep{Chon2012, Duffy2018}, making them significantly cooler than the background regions. The background regions defined in this work have a significantly higher temperature and thus can be assumed to be part of the cocoon shock, rather than the rib-like structures. Table \ref{tab:therm} does show a slightly worse fit quality regions B3 and B4, which could indicate some mixing with enriched gas from the core.  However, the reduced $\chi^2$ values in these regions, $1.07$ for B3 $1.03$ for B4, show that these are still acceptable fits.

\subsection{Non-thermal emission components}
\label{results_postdib}

\begin{figure*}
   \includegraphics[width=1.0\textwidth]{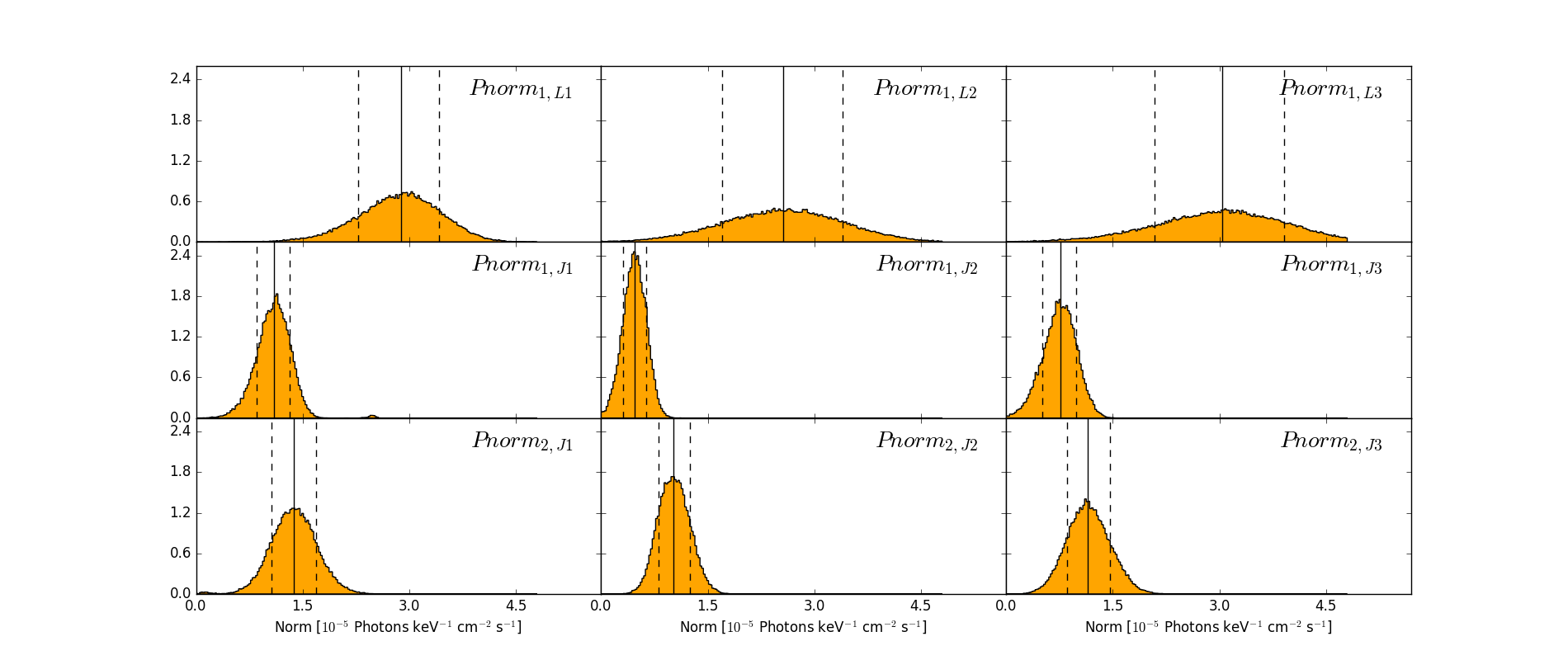}
    \includegraphics[width=1.0\textwidth]{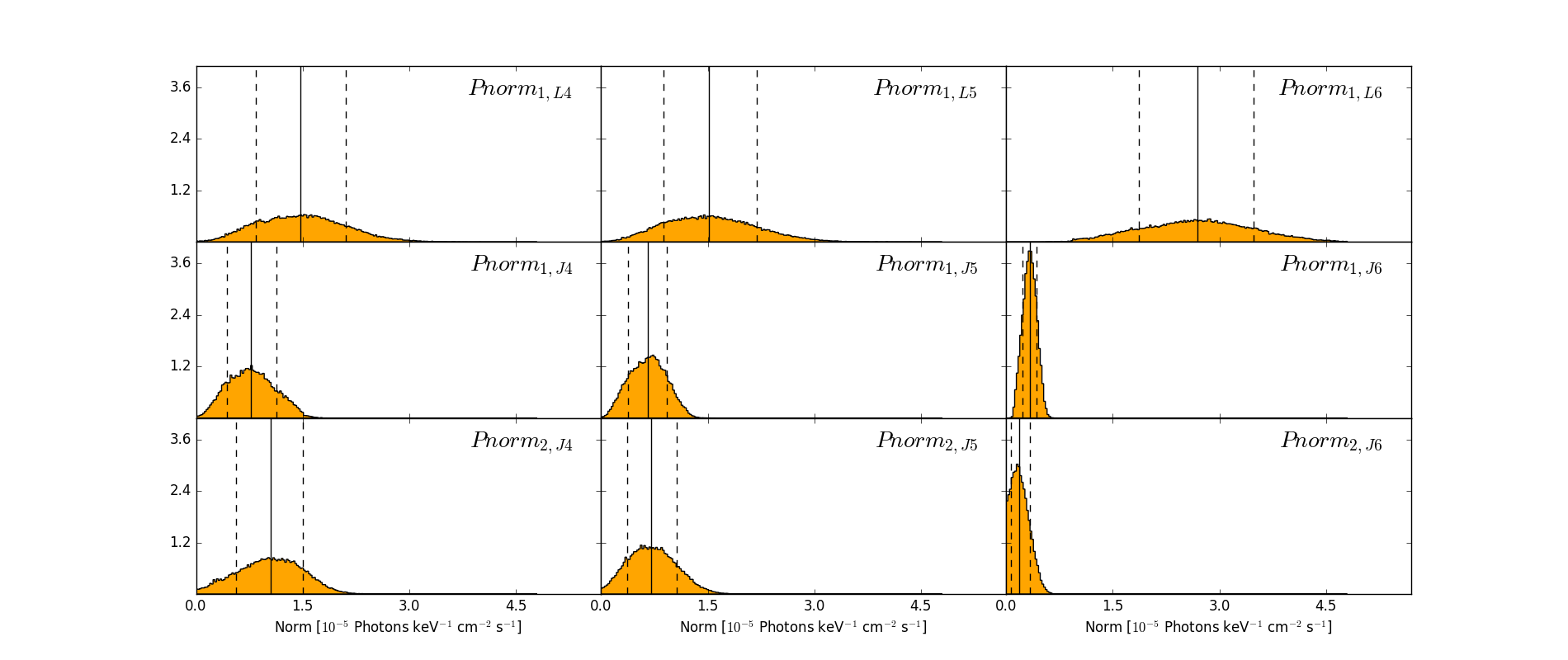}

  \caption{Top: Posterior distributions for the power law normalisations in the eastern lobe. Normalisations indicate the photon flux density at 1 keV. The top row shows the non-thermal emission in each lobe region. The middle row shows the non-thermal lobe emission in each jet region, and the bottom row shows the non-thermal jet emission each jet region. The solid lines indicate the median, the dashed lines the 14th and 86th percentiles.
  Bottom: As above, for the western lobe.} 
    \label{fig:plnorm}

\end{figure*}

\begin{figure}
   \includegraphics[width=0.45\textwidth]{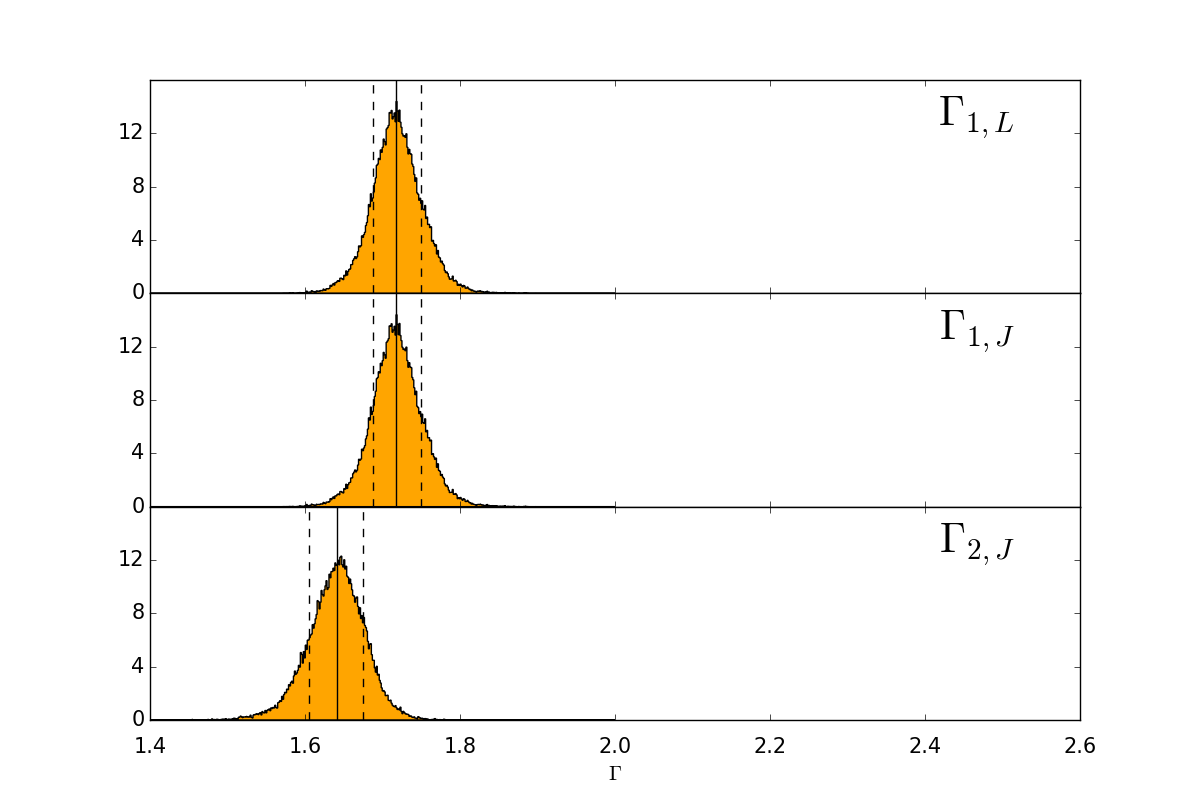}
   \includegraphics[width=0.45\textwidth]{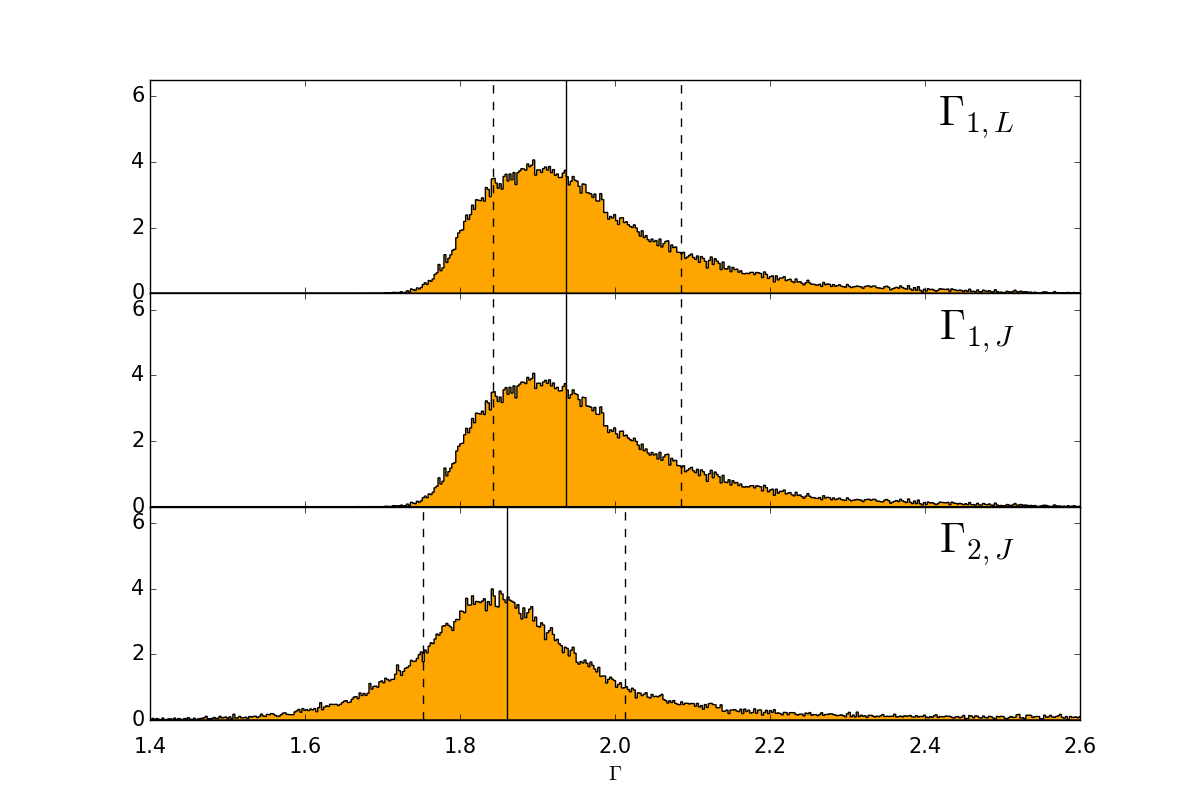}

  \caption{Top: Posterior distributions for the power law photon indices in the eastern lobe. The top figure shows the photon index of the non-thermal emission in the lobe regions. The middle figure shows the photon index of the non-thermal lobe emission in the jet regions. The bottom figure shows the photon index of the non-thermal jet emission in the jet regions. The solid lines indicate the median, the dashed lines the 14th and 86th percentiles. Bottom: As above, for the western lobe} 
    \label{fig:plgamma}

\end{figure}
  
 \begin{table}
 \centering
{\renewcommand{\arraystretch}{1.2}
\caption{Flux density and power law of the non-thermal emission from the jets and lobes.  }
\begin{tabular}{ccccc}
\hline \hline
&& S$_{\rm 1 keV} $ \textsuperscript{a} (nJy) & $\Gamma$ &  \\ \hline
\textbf{East} & Lobe & $71_{-10}^{+10}$  & $1.72_{-0.03}^{+0.03}$  \\
 & Jet & $24_{-4}^{+4}$&   $1.64_{-0.04}^{+0.04}$ \\ \hline
\textbf{West} &  Lobe & $50_{-13}^{+12}$ & $1.97_{-0.10}^{+0.23}$  \\
 & Jet  & $13_{-5}^{+4}$&  $1.86_{-0.12}^{+0.18}$ \\ \hline
 \label{tab:flux_results}
\end{tabular}}

 \raggedright{\setlength\parindent{4em} \textsuperscript{a} Flux density at 1 keV }

\end{table}

From the model comparison we conclude that there is non-thermal emission from both the lobes and the X-ray jets. This means that in each set of lobe and jet regions, there are 3 non-thermal components and 3 associated normalisations: $Pnorm_{1,L}$ for the normalisation of the lobe emission in the lobe region, $Pnorm_{1,J}$ for the normalisation of the lobe emission in the jet region, and $Pnorm_{2,J}$ for the normalisation of the jet emission. Correspondingly, there are also 3 photon indices: $\Gamma_{1,L}$ for the lobe emission in the lobe regions, $\Gamma_{1,J}$ for the lobe emission in the jet regions, and $\Gamma_{2,J}$ for the jet emission. Because we used the posterior distributions from the lobe regions as a prior to constrain the lobe emission in the jet region, we expect these posterior distributions to look similar. We comment further on this in section \ref{sec:disc:disent}.

We show the posterior distributions of the non-thermal emission components from the lobes and jets in Figs. \ref{fig:plnorm} and \ref{fig:plgamma}. We list the total flux and the photon index of each lobe and each jet in Table \ref{tab:flux_results}. We note that the lobe flux densities on the eastern and western side are consistent with those found by \cite{Yaji2010} and \cite{Hardcastle2010}. For plots showing the correlation between the normalisation and photon index of each component in each region, we refer to Appendix \ref{appendix:corrs}.

We compared the photon indices obtained from the posterior distributions to the radio spectral index. \cite{Spinrad1985} finds an average low-frequency spectral index $\alpha$ of 0.74 for the lobes. This agrees well with our photon index of $1.72_{-0.03}^{+0.03}$ in the eastern lobe, but not with the value of $1.97_{-0.10}^{+0.23}$ in the western lobe. We discuss possible causes for the differences between the lobes in section \ref{sec:disc:difs}.

\subsection{Non-thermal pressure in the lobes and X-ray jets}
\label{sec:results:pressures}
\begin{table}
\centering
\caption{Comparison of the rim lobe pressures to the modeled IC lobe pressures, for $\kappa=0$.}
\begin{tabular}{p{0.8cm}p{1.9cm}p{0.7cm}p{1.9cm}p{1.1cm}}
\hline \hline
& $p_{\rm rim}$\textsuperscript{a} \newline ($10^{-10}$ erg cm$^{-3}$) & $\gamma_{\rm min}$ \textsuperscript{b}  & $p_{\rm IC}$ \textsuperscript{c} \newline ($10^{-10}$ erg cm$^{-3}$) & B ($\mu$G) \textsuperscript{d} \\ \hline
\textbf{East} & $10.4 \pm 0.4 $  & 1 & $5.8_{-1.4}^{+2.0}$ & $42_{-3}^{+3}$  \\
 &  & 10 & $2.2_{-0.4}^{+0.4} $& $42_{-3}^{+3}$ \\ \hline
\textbf{West} & $8.4 \pm 0.2$ & 1 & $140_{-116}^{+1690}$ & $45_{-5}^{+15}$   \\
 &   & 10 &  $18_{-10}^{+40}$ & $45_{-5}^{+15}$\\ \hline
\label{tab:lobe_P}
\end{tabular}
 \raggedright{\setlength\parindent{0em} \textsuperscript{a} The rim pressures for the eastern and western lobe, taken from \cite{Snios2018}. \\
 \textsuperscript{b} The lower cutoff to the electron energy distribution. \\
 \textsuperscript{c} The IC pressures, obtained from \textit{synch}. See text for details. \\
 \textsuperscript{d} The magnetic field strength, obtained from \textit{synch}. }

\end{table}

\begin{table}
\caption{The IC relic jet pressures, for $\kappa=0$. }
\begin{tabular}{p{0.8cm}p{1cm}p{2cm}p{1.5cm}}
\hline \hline
&   $\gamma_{\rm min}$  \textsuperscript{a}  & $p_{\rm IC}$  \textsuperscript{b} \newline ($10^{-10}$ erg cm$^{-3}$) & B ($\mu$G) \textsuperscript{c}\\ \hline
\textbf{East}  & 1 & $7.9_{-3.3}^{+5.8}$ & $27_{-4}^{+5}$  \\
 &  10 & $4.0_{-1.3}^{+2.6} $&$27_{-4}^{+5}$ \\ \hline
\textbf{West} & 1 & $99_{-76}^{+540}$ & $17_{-3}^{+7}$   \\
 &   10 &  $22_{-12}^{+40}$ & $17_{-3}^{+7}$\\ \hline
\label{tab:jet_P}
\end{tabular} 

 \raggedright{\setlength\parindent{0em}  \textsuperscript{a} The lower cutoff to the electron energy distribution. \\
 \textsuperscript{b} The IC pressures, obtained from \textit{synch}. See text for details. \\
 \textsuperscript{c} The magnetic field strength, obtained from \textit{synch}. }
\end{table}

\subsubsection{The lobes}
\label{sec:lobepressure}

The parameters for the non-thermal emission that we find under model $M_{\rm L1}$ and $M_{\rm J1}$ were used to model the pressure in the lobes. We compared the pressures found from the models with the rim pressures as calculated by \cite{Snios2018}.  These pressures are determined from X-ray spectra of compressed gas in regions between the cocoon shock and the lobes. The average rim pressures for the eastern and western lobe are $p_{\rm rim, east} = (10.4 \pm 0.4) \times 10^{-10} $ erg cm$^{-3}$  and $p_{\rm rim,west} = (8.4 \pm 0.2) \times 10^{-10}$ erg cm$^{-3}$.  By comparing the rim pressures with the non-thermal pressures from our models we are able to constrain lobe parameters such as the fraction of non-radiating particles and the lower limit to the electron energy distribution, denoted as $\kappa$ and $\gamma_{\rm min}$ respectively. 

The IC pressures were modeled with the inverse-Compton code of  \cite{Hardcastle1998}, called \textit{synch}. The code takes into account both IC/CMB as well as SSC, which is an important component in the Cyg A lobes \citep{Hardcastle2010, Yaji2010}.  The code calculates the total energy density in a certain volume, including the energy density from both particles and the magnetic field. We have modeled the electron distributions as broken power laws with an age break.

We calculated the volume of each lobe by approximating them as a capped ellipsoid, inclined to the line of sight with an angle of 55 degrees \citep{Vestergaard1993}, based on the lobe region sizes. We obtain total volumes of $6.8 \times 10^{68}$ cm$^{3}$ for the eastern lobe, and $7.4 \times 10^{68}$ cm$^{3}$ for the western lobe. 

We normalised the synchrotron spectrum to the flux inside the lobe and jet regions on the 4.5 GHz VLA radio map \citep{Carilli1991}. Because no radio brightness enhancement is observed at the location of the X-ray jet, we assume that all of the radio emission in the lobe and jet regions can be attributed to the lobes. We find flux densities 211 Jy for the eastern lobe and 156 Jy for the western lobe. 

The break frequency $\nu_B$ varies over a range from 1 to 10 GHz in the lobes of Cyg A \citep{Carilli1991}. We have modeled each lobe with a single average break frequency of 5GHz. We have assumed that the photon index increases by 0.5 beyond the break frequency.  Initial runs with \textit{synch} show that the magnetic field in the lobes is around $\sim 40$ $\mu$G. This translates to electron Lorentz factors of  $\gamma_B  \sim 7000$ at the break frequency.

The choice for the lower cutoff to the electron energy distribution,  $\gamma_{\rm min}$, can significantly affect the calculated pressure. The higher the photon index, the steeper the slope, and the more the low-energy electrons contribute to the total pressure. The value of $\gamma_{\rm min}$ is unknown, we calculate the pressures for $\gamma_{\rm min} =1$ and $\gamma_{\rm min} = 10$ . The upper cutoff to the electron distribution was set at  $\gamma_{\rm max} = 10^5$, giving a cutoff in the synchrotron spectrum at $\simeq 10^{12} $ Hz. 

The slope of the electron energy distribution, $p$, is directly related to the photon index by $p = 2 \Gamma -1 $. We have used  $\Gamma$ from our models to determine $p$. We now have assumptions for $\gamma_{\rm min}$ and $\gamma_{\rm max}$, the photon indices and normalisations of the IC spectrum as determined from the posterior distributions, and the normalisation and $\nu_B$ of the synchrotron spectrum determined from the VLA radio data. With these,  the magnetic field strength and energy density in the lobe can be modeled with \textit{synch}.  

We initially assumed an equipartition magnetic field and determined what the predicted X-ray flux would be in this case, using the median value of the photon index posterior distribution. Assuming $\gamma_{\rm min}=1$, we find equipartition fields of 95 $\mu$G and 210 $\mu$G in the eastern and western lobe respectively.  For $\gamma_{\rm min}=10$, the equipartition fields are 73 $\mu$G and 130 $\mu$G. However, for both values of $\gamma_{\rm min}$ the equipartition field underpredicts the observed X-ray flux by factors of a few, implying that the true magnetic field is below the equipartition value.

We then modeled the lobe pressure and magnetic field strength by using the observed X-ray flux from the posterior distributions.  For each lobe, we took 300 random samples of the non-thermal normalisation and photon index, and ran $\textit{synch}$ for each of these parameter sets. The resulting distributions of the model  pressure for each lobe, and a comparison with the rim pressures, are shown in Table \ref{tab:lobe_P}.

In the calculations above we have assuming that the fraction of non-radiating particles in the lobe, $\kappa$, is zero. Because the magnetic field strength is below equipartition, the total energy is dominated by the particle energy. We can thus assume the IC pressures to scale linearly with $\kappa$ + 1. 

As was previously reported by \cite{Hardcastle2010}, we find that the non-thermal lobe flux in both lobes is dominated by SSC. In the eastern lobe, SSC makes up about 80\% of the total non-thermal flux. In the western lobe, the ratio spread is wider because of the wider distribution of $\Gamma$, but SSC makes up about 50-90\% of the non-thermal flux.

\subsubsection{The X-ray jets}
\label{sec:jetpressure}

We have modeled the X-ray jets as an inverse-Compton emitting population of electrons, with \textit{synch}. This corresponds to the IC relic jet model proposed by \cite{SB2008}. The energy density and pressure in the X-ray jet can be calculated in the same manner as the lobes and compared to the lobe and rim pressures on each side. 

For the volume, we have used the defined jet regions and assumed that they are tubular in shape. We also assume an inclination angle to the line of sight of 55 degrees. This yields total volumes of  $3.8 \times 10^{67}$ cm$^{3}$ and $4.2 \times 10^{67}$ cm$^{3}$ for the eastern and western jets respectively.

It is difficult to model the synchrotron spectrum from the X-ray jets, because little to no emission is observed from these features. In the IC relic jet model,  the adiabatic expansion of the jet should cause the Lorentz factors of the electron population to go down. The adiabatic expansion combined with synchrotron aging caused the relic to have faded beyond detection at radio wavelengths. 

We have looked for evidence of the relic jet in the LOFAR 150 MHz data \citep{McKean2016}. In the eastern lobe, we observe an enhancement in the brightness and spectral index map at roughly the location of the relic jet. However, the enhancement seems to be between two of the X-ray jet knots. In the western lobe, there is a slight brightness enhancement as well, although the corresponding spectral index enhancement is weaker than in the eastern lobe. Similarly, the enhancement is on the path of the relic jet, but not at the same location as the X-ray jet knot. In both lobes, there seems to be a faint brightness enhancement that is roughly aligned with the path of the relic jet. These brightness and spectral index enhancements, although weak and not well aligned, provide a hint that the X-ray jets are non-thermal in origin, consistent with our results from the model comparison.

Regardless of whether the radio features seen in the LOFAR maps are associated with the X-ray jet, it is difficult to determine what the radio spectrum of the IC relic jet would look like. We have used the LOFAR radio map to set an upper limit to the number density of electrons in the relic jet plasma. At most, the radio emission per unit volume of the relic jet on the 150 MHz LOFAR map cannot be more than that of the lobe. Using this assumption, we obtain maximum flux densities of 220 Jy at 150 MHz for the eastern jet, and 155 Jy at 150 MHz for the western jet. We note that this upper limit to the radio flux effectively corresponds to a lower limit on the modeled pressures. The lower the radio flux, the further below equipartition the relic jet will be, and the higher the modeled IC pressure.

While the relic jet only generates a relatively small number of synchrotron photons, it is embedded inside the lobe and subjected to its synchrotron photon field. We therefore also considered a third IC component, which is the external Compton of the lobe and hotspot photon fields passing through the jet. We modeled the spectrum of the lobe by assuming that the radio emission mechanism is isotropic, and we assumed that the emission coefficient $j_{\nu}$, is a constant throughout the lobe. The average intensity at a specific wavelength is then 
\begin{equation}
\label{eq:avg_J}
J_{\nu} = j_{\nu} \int{\frac{dV}{4 \pi \ell^2}},
\end{equation}
where V indicates the volume, $\ell$ the path along a ray from the source to the region of interest, and $J_\nu$ the average intensity. We assume axial symmetry for the lobe, and use cylindrical polar coordinates $r, \phi, z$. Then, integrating over the angle $\phi$ yields
\begin{equation}
\label{eq:cyl_integral}
J_{\nu} = j_{\nu} \int{ \frac{ r \, dr\, dz }{2 \sqrt{\{z^2 + (r - x_0)^2\}\{z^2 + (r + x_0)^2\}}}} \, ,
\end{equation}
where $x_0$ is the distance from the axis of the point of interest in the plane $z = 0$. We evaluate the intensity at the centre of the jet ($x_0 = 0$).  We also assume the jet is a cylindrical tube inside the lobe, and so we integrate over the cylinder radius $r$ between $r_{j}(z)$ and $r_{l}(z)$.  This reduces the integral to a 1-dimensional integral over $z$:
\begin{equation}
\label{eq:cyl_integral2}
J_{\nu} = \frac{j_{\nu}}{4} \int{  \ln{\Big( \frac{z^2 + r_l^2}{z^2 + r_j^2}}\Big) dz}.
\end{equation}
The radius of the lobe and the jet, $r_j$ and $r_l$, are both functions of $z$. We have approximated both functions for each lobe by manually measuring the radius at several points along the z-axis and linearly interpolating between these points. 

We used equation \ref{eq:cyl_integral2} along with the radio flux from the VLA data, to calculate the average intensity at 4.5 GHz. We then modeled the spectrum of the lobe as a broken power law, using a break frequency of 5GHz. The spectral index of the lobe model is drawn directly from the posterior distributions and therefore varies from sample to sample.

Additionally, we estimated the influence from the hotspots by modelling them as point sources. We again took the radio flux of the hotspots from the VLA data, finding flux densities of 117 Jy for the eastern hotspots and 152 Jy for the western hotspots. We then calculated the average flux between the minimum and maximum hotspot - jet distance. We modeled the hotspot spectrum as a broken power law, with a break frequency at 10GHz and a photon index of 1.5. 

Because the spectrum of the relic jet is unknown, an assumption for the break frequency has to be made.  If the break frequency in the radio spectrum is too low, the IC spectrum would have a turnover below the 0.5 - 7.0 keV range, and the photon index that we see in the X-ray data would be the photon index beyond the turnover. We consider this unlikely, especially in the eastern X-ray jet, as the slope before the turnover would then be as flat as $\sim 1.1$. 
This places constraints on how low the break frequency can be. By modelling the jets with \textit{synch}, we find that the break frequency should not be lower than $\sim 4 $ GHz. Below that value, Compton scattering of synchrotron photons originating from the lobes, the dominant component in the X-ray jet flux, starts to turn over enough that it noticeably affects the slope of the total IC spectrum. We therefore place the break frequency at this value. The break frequency of 4 GHz and a magnetic field strength of $30$ $\mu$G correspond to an electron break Lorentz factor$\gamma_{B} \sim 7000$.  The same cutoffs and lower and upper limits were applied as for the lobes: the mininum frequency for the synchrotron spectrum is 1 MHz. The jet pressures were calculated for $\gamma_{\rm min}=1$ and $\gamma_{\rm min} = 10$. The maximum electron Lorentz factor was set at $\gamma_{\rm max} = 10^5$ , which gives a cutoff in the synchrotron spectrum at $\sim 10^{13}$ Hz. 

We calculated the distribution of model pressures the same way as for the lobes: we took $300$ random samples from the posterior distributions of the jet normalisation and photon index. We assumed $\kappa=0$ and we calculate pressures for both $\gamma_{\rm min}=1$ and $\gamma_{\rm min}=10$. The results are listed in Table \ref{tab:jet_P}. We find that external Compton scattering dominates the total flux. In the eastern jet, we find that the external Compton from the lobe photons contributes approximately $65\%$ to the total flux, SSC of the jet photons approximately $15\%$, and IC/CMB scattering $20\%$. In the western jet, the wider distribution of $\Gamma$ causes a greater spread in these fractions. External Compton of the lobe photons contributes $40-60\%$, SSC of the jet photons $5-30\%$ and IC/CMB scattering $5-40\%$. This seems  shows that the relic jet is not purely an IC/CMB X-ray source, but that SSC needs  to be taken into account in an IC relic jet model.

\section{Discussion }
\label{sec:discussion}

\subsection{Disentangling the lobe and jet emission components}
\label{sec:disc:disent}
 
The fact that the model in the jet regions contains two separate power laws with similar photon indices, means that degeneracies are a concern. It is possible that the MCMC routine in the jet regions does not manage to fully disentangle the lobe emission from the jet emission. We have attempted to minimise this problem by making the amount of lobe emission in the lobe regions a prior for the amount of lobe emission in the jet regions. By assuming that the surface brightness of the non-thermal lobe emission is constant, the MCMC routine has less difficulty separating the non-thermal emission in the jets into two different components. 

From the middle and bottom rows of both panels of Fig. \ref{fig:plnorm}, it appears that the two power laws can be distinguished from one another in every region. If they were not, we would expect to see most of the non-thermal emission attributed to just one of the power laws and the other posterior distribution approaching zero. Instead, the results show distinct unimodal peaks for the lobe component and the jet component in each jet region.
 
Because we used the posterior distributions of $Pnorm_{1,L}$ and $\Gamma_{1,L}$ from the lobe model as priors for the parameters $Pnorm_{1,J}$ and $\Gamma_{1,J}$ in the jet model, we expect the posterior distributions of the corresponding jet and lobe parameters to be very similar. We compared the prior distributions to test this, and find that the median and the spread of each distribution agrees to a few \% precision. This indicates that the MCMC routine for the jet regions did not stray far from the prior distributions given by the MCMC sampling of the lobe regions. Our assumption of constant surface brightness between a lobe region and its corresponding jet region seems therefore to be reasonable. 

We investigated whether the ratio of jet to lobe flux on each side, obtained from the posterior distributions, agrees with the jet to lobe count ratio from the event data. The count ratio is determined from the event files as follows:  we determine the number of counts in the jet region. We then subtract the number of counts in the lobe region, scaled to the area of the jet region. We then assume that 50-70\% of the counts in the lobe region are thermal, and subtract this number from both the lobe and the jet. We are then left with estimates for the number of non-thermal counts in the lobe and the jet. In the eastern lobe, we find a flux ratio of  $1.5_{-0.4}^{+0.6}$, and a count ratio of $2.0 - 2.7$. On the western side, we find a flux ratio of $ 1.1_{-0.6}^{+0.9}$ and a count ratio of $1.7 - 2.1$.  

The count ratio is slightly higher than the flux ratio on both sides, which raises the possibility that our model underestimates the jet flux.  However, the higher count ratio translates to  only a modest difference in the jet flux. In the eastern jet, a count ratio of $2.0 - 2.7$ corresponds to a jet flux of 26 - 28 nJy, while in the western jet a count ratio of $1.7 - 2.1$ corresponds to a jet flux of 15 - 17 nJy. Both estimates are within the errors of the jet flux distribution from the model. The estimate provided by the count ratio seems to agree well with the flux ratio obtained from the model.

\subsection{A two-temperature thermal model for the ICM}
\label{sec:disc:twoT}

As shown in section \ref{sec:results:thermal}, there is a significant difference between the ICM on the eastern and the western side of Cyg A. The temperature increase on the western side is in the direction of the merger with nearby subcluster Cyg NW, and roughly corresponding to the direction of the outburst.  The fact that the temperature increase is only on one side would suggest that a shock created by the merger is the underlying cause of the temperature increase. Moreover, it is possible that the merger shock has enhanced features that were already there, perhaps imprints of previous cycles of AGN activity, or features created by sloshing motions. For a more extended discussion of the complex merger region, we refer to Wise et al. (in prep.).

Regardless of the cause of the temperature difference between east and west, there is reasonable cause to suspect that the ICM surrounding the western cocoon shock may actually be better described by a  two-temperature thermal plasma.  For example, one could imagine a geometry where the ICM shocked by the merger is a layer of hot $ \sim 10$ keV  material  that is partly projected in front of the lobe, while the underlying ICM has a temperature of $\sim 6$ keV, the same as on the eastern side. 

\cite{Mazzotta2004} have investigated the effect of fitting a single-temperature thermal model to a two-temperature plasma with \textit{Chandra}.  For high gas temperatures ( $ > 5 $ keV, and low abundances ($ < 1.0 $ Z$_\odot$),  they find that a single-temperature thermal model fit is often statistically indistinguishable from a fit with a two-temperature thermal model. This is because when the gas temperature is high, the gas will be more highly ionised, making it more difficult to distinguish spectra with differing temperatures. In the case of Cyg A's western lobe, assuming two temperature components of $\sim 6 $ and $\sim 10$ keV, and an abundance of $\sim 0.5$ Z$_\odot$, the results from \cite{Mazzotta2004} indicate that the emission from this plasma would be very well fit by a single-temperature thermal model. Although the structure of the hot plasma around the western lobe might be complex, we therefore expect that using a single-temperature thermal model provides an adequate enough description of the spectral data.

\subsection{Difference between the eastern and western lobes}
\label{sec:disc:difs}

The posterior distributions show clear differences between the eastern and western side, with the western side being both fainter and having a steeper X-ray spectrum. The steeper X-ray spectrum maps onto a steeper electron spectrum with a larger fraction of electron energies with low $\gamma$. This translates into higher energy densities and pressures on the western side. 

The photon index on the eastern side agrees well with the value of \cite{Spinrad1985}, as well as with the spectral index obtained from the LOFAR data by \cite{McKean2016}. All of these show average spectral indices $ \alpha \sim 0.7$.  

If the photon index in the western lobe accurately reflects the average photon index, we suggest that either aging or adiabatic losses may have caused a turnover in the IC spectrum somewhere below 7.0 keV. This would explain why the photon index is higher and less constrained. It would also mean the pressures calculated by \textit{synch} are overestimated on the western side, because the low $\gamma$ range of the electron energy spectrum would have a lower photon index than we have modeled. 

We looked at the available VLA and LOFAR data to see if there are differences between the lobes in the radio. In both radio maps, we find that the western lobe is roughly 30\% fainter than the eastern lobe. However, there appear to be no appreciable differences between the lobes in terms of spectral index and break frequency \citep{Perley1984,McKean2016}.

\cite{Snios2018} estimate that the total volume of the western lobe is about 40\% larger than the eastern lobe. Their estimate includes the volume of the shocked cocoon not just of the lobes. We note that in our own estimate of the lobe volume, the western lobe is only about 10\% larger than the eastern lobe. However, the lobe regions that we have defined do not exactly follow the radio lobe, and also have the hotspot regions cut out. Therefore, the volume calculated from the lobe regions is not necessarily accurate for the lobe as a whole.

If the western lobe is indeed bigger than the eastern lobe by a few tens of percent, then this could indicate additional adiabatic expansion on the western side, which would reduce both the magnetic field strength and the particle energies. Under simple assumptions for adiabatic expansion, $B \propto V^{-2/3}$and $\gamma \propto V^{-1/3} $, which means we would expect the magnetic field strength in the western lobe to be 80\% of that in the eastern lobe. The magnetic field strengths in Table \ref{tab:lobe_P} do not differ significantly between east and west, although the errors on the western side are large. 

For a synchrotron spectrum, the Lorentz factor of an emitting electron and the emitted frequency are related as
\begin{equation}
\nu \simeq \frac{\gamma ^2 q B } { 2 \pi m_e}\, , 
\end{equation}
while in the corresponding SSC spectrum, the Lorentz factor is related to the energy as
\begin{equation}
E =\frac{\gamma^4 \hbar q B} {m_e} \, .
\label{eq:ESSC}
\end{equation}
Equation \ref{eq:ESSC} and the scaling relations for $B$ and $\gamma$ imply that the characteristic energy scales with the volume as $E \propto V^{-2}$. If the western lobe is 40\% bigger, the turnover in the IC spectrum of the western lobe could therefore be at 50\% of the energy of that of the eastern lobe.

A break frequency $\nu_B$ at $5$ GHz and a magnetic field strength of around $40$ $\mu G$, yield break Lorentz factor $\gamma_{B} \sim 7000$, and a break energy $E_B \sim 1$ keV. This means that a turnover of the SSC part of the spectrum would be a plausible explanation for the steeper spectrum in the western lobe, especially because SSC makes up a significant amount of the total non-thermal flux in the lobes. To look for evidence of a spectral turnover, we re-evaluated the lobe spectra in two separate energy bands of 0.5 - 2.0 keV and 2.0 - 7.0 keV. We repeated the MCMC analysis of model $M_{\rm L1}$ in both lobes, allowing each energy band to have a different photon index. In the eastern lobe, we find $\Gamma_{0.5-2.0} =  1.56_{-0.07}^{+0.07}$ and $\Gamma_{2.0-7.0} = 1.72_{-0.04}^{+0.04}$. In the western lobe, we find $\Gamma_{0.5-2.0} = 1.80_{-0.11}^{+0.19}$ and $\Gamma_{2.0-7.0} = 1.94_{-0.10}^{+0.25}$. While the errors in the western lobe are too large to distinguish between these photon indices with any statistical certainty,  the numbers are consistent with the possibility that we are indeed seeing a turnover in the non-thermal X-ray spectrum of the western lobe. Somewhat more surprising is that a similar effect is also observed in the eastern lobe, given that the photon index of $1.72_{-0.03}^{+0.03}$ agrees well with the radio data, and we do not expect the average photon index to be significantly lower than this.  However, we note that our models are limited by the fact that each lobe and jet are modeled with only a single photon index. Therefore, the variation seen between the low and high energy bands on both sides should be taken as a sign that the true electron energy distributions are more complicated than assumed here.

The correlation plots in Appendix \ref{appendix:corrs} provide further insight as to why the photon index in the west is higher. In the eastern lobe, the model shows a non-thermal surface brightness in region L1 that is clearly larger than in regions L2 and L3. This is consistent with the radio maps, which show more continuum emission closer towards the hotspots. By contrast, in the western lobe, L6 is not significantly brighter than regions L4 and L5. If the X-ray photon index follows the radio spectral index, we would expect it to be significantly higher closer towards the AGN. Therefore, the reason that the photon index in the western lobe is higher could be explained by the fact that L4 and L5 contribute more to the total flux in the west than L2 and L3 do in the east. Because these inner regions have a higher photon index, the average photon index for the entire lobe will also be higher. 

This raises the question, what has caused the underlying discrepancy?. As we have discussed at the beginning of this section, it is possible that the environment in the lobes and jets itself is different, and that the turnover in the western side is at lower energies than on the eastern side. Additionally, \cite{Snios2018} have shown that the shock on the eastern side is stronger, and so the shock could have managed to create more energetic particles in region L1, pushing the average photon index in the eastern lobe to a lower value.  A third possibility is that the geometrical assumptions that we have made have influenced the results. In particular, the assumptions that link the normalisations of emission components between corresponding background, lobe and jet regions might be inaccurate, perhaps because the western lobe is less symmetric than the eastern lobe. This would cause too much or too little emission to be attributed to one of the emission components. 

\subsection{Contribution of infrared photons to the IC flux}

In our analysis of the we have considered synchrotron and CMB photons as seed photons for the IC process. However, infrared photons emitted from the AGN and the dust around the AGN can additionally provide a significant contribution to the total IC flux \citep{Brunetti1997}. In this section, we estimate approximately how much these photons would contribute to the IC flux.

\cite{Weedman2012} calculate the strength of several IR spectral features in Cyg A, but do not provide a value for the total infrared luminosity $L_{\rm IR}$. However, we have made use of the scaling relationship between $\nu L_{\nu}(7.8 \, \mu m)$ and $L_{IR}$, which is $\log[ L_{\rm IR} / \nu L_{\nu}(7.8 \, \mu m)] = 0.80 \pm 0.25 $ in AGN with silicate absorption \citep{Sargsyan2011}. Using the value of  $f_{\nu}(7.8 \, \mu m) = 54$ mJy from \cite{Weedman2012} and using the upper limit of the scaling relationship, we find $L_{\rm IR} \sim 1.8 \times 10^{45}$ erg s$^{-1}$.  This is consistent with the scaling relationship between bolometric luminosity and infrared luminosity found in that same paper,  using  $L_{\rm bol} \sim 3.8 \times 10^{45}$ erg s$^{-1}$ from \cite{Privon2012}.

We calculated the number density of infrared photons in the lobe as $N_{\rm IR} = L_{\rm IR}/(4 \pi c d^2 E_{\rm IR})$, where we have used $d = 40 $ kpc as an average distance in the middle of the lobe, and $E_{\rm IR} = 3.3 \times 10^{-14}$ erg ($\lambda_{\rm IR} = 60 $ $\mu$m ) as the characteristic energy of an infrared photon, corresponding to the peak of the SED \citep[][]{Privon2012}. This yields $N_{\rm IR} = 10 $ cm$^{-3}$.  Meanwhile, the energy density of the CMB is $u_{\rm CMB} =  4.1 \times 10^{-13} (1 + z)^{4}$ erg cm$^{-3}$ \citep{Harris1979}, yielding a photon number density $N_{\rm CMB} = 360$ cm$^{-3}$.

While the number density of CMB photons is larger, we also have to take into account  that infrared photons have higher energy and therefore have access to a larger number of electrons to be inverse-Compton scattered to keV energies. A 60 $\mu$m photon requires $\gamma \sim 200$ to be upscattered to 1 keV, while a CMB photon requires $\gamma \sim 1000$ to be upscattered to 1 keV.  Using the slope of the electron spectrum $p=2.4$ in the eastern lobe, we find a relative electron number density of $N_{\gamma_{1000}} / N_{\gamma_{200}} = 0.11$. For the flux ratio at 1 keV of IC/CMB flux to infrared IC flux, we then estimate $ f_{\rm CMB} / f_{\rm IR} \sim $ $ N_{\rm CMB} N_{\gamma_{1000}}  / \, N_{\rm IR} N_{\gamma_{200}}  = 4$. 

The modelling in section \ref{sec:results:pressures} has shown that synchrotron radiation is the dominant component in both lobes, and that the CMB makes up between $10-50\%$ of the flux. If the IC/CMB flux is 4 times higher than the infrared IC flux, as our estimate indicates, the infrared IC flux would contribute 2.5 - 12.5\% to the total IC flux.

We expect that including the infrared spectrum as an additional photon field in our model would reduce the pressures, as fewer electrons would be needed to produce the same X-ray flux.  However, since the maximum contributions of the infrared photons are less than the errors in the pressures, including them in the model would not have significantly altered our results.

\subsection{The X-ray jets as IC relic jets}

MCMC sampling of the Bayesian models for the jets and lobes of Cyg A has yielded good constraints on their photon indices and flux densities. However, turning these into pressures introduces additional errors. The pressure is strongly dependent on the photon index in particular, which leads to very large uncertainties, particularly in the western lobe. Additionally, it is not known what the radio spectrum of the relic jet would be. This has forced us to make assumptions about the shape of the spectrum. 

The magnetic field strengths found in Table \ref{tab:lobe_P} are a factor of 2-6 below equipartition. This seems to be a typical value for FRII radio galaxies \citep{Croston2005, Ineson2017}. We compare the rim pressures with the IC pressure listed in Table \ref{tab:lobe_P} to constrain the particle content $\kappa$ in the lobes. In the eastern lobe, the assumption of $\kappa=0$ yields a pressure that is inconsistent with the rim pressure. Depending on the choice of $\gamma_{\rm min}$, we require $ 1 < \kappa < 5 $ in the eastern lobe to match the rim pressures.  \cite{Croston2014} developed a model for FRI radio galaxies where jet entrainment of protons/ions could provide the necessary additional pressure. A similar process could also be at work here. 

On the eastern side, if $\kappa$ and $\gamma_{\rm min}$ in the jet have the same value as in the lobe, the lobe and jet pressure are the same within the errors.  However, it is unclear what determines the particle content of the jet and how it is related to the particle content in the lobe. It is possible that $\kappa$ in the jet is lower or higher than in the lobe. On the western side, the calculated jet and lobe pressures are the same within the errors as well, assuming the same $\kappa$ and $\gamma_{\rm min}$. At $\kappa=0$ and  $\gamma_{\rm min}=1$, both of these pressures are much higher than the western rim pressure.  This would imply that $\kappa=0$ and that $\gamma_{\rm min}$ is larger than in the eastern side. However, it is difficult to imagine a scenario where $\kappa$ is significantly different between the eastern and western side. Given the discrepancy in photon indices, and the possibility of a turnover in the western side due to adiabatic losses or aging effects, we consider it more likely that we have overestimated the western lobe and jet pressures, as discussed in section \ref{sec:disc:difs}.

Another possibility is that $\gamma_{\rm min}$ is higher than assumed in our analysis. A consequence of the steeper spectrum on the western side is that the pressure falls off more quickly with increasing $\gamma_{\rm min}$. Therefore, if  $\gamma_{\rm min}$ is higher on the western side than on the eastern side, it is possible to make the western lobe pressure consistent with the eastern lobe pressure. 

We note that our relic jet model is significantly different from the pure IC/CMB jet with very low magnetic field, proposed by \cite{SB2008}. As already noted in Section \ref{sec:intro}, a problem of the IC/CMB jet model is that it requires a significantly higher electron density, which would result in much higher electron pressures in the jet. In our model, the magnetic field in the X-ray jet is lower than in the lobes by a factor of $\sim 35-65 \%$. This allows for a higher electron density in the jet while still producing less synchrotron radiation than the lobe. At the same time, the magnetic field is still high enough that the produced synchrotron photons are energetic enough to be IC scattered to the keV energy range. In this way, SSC can be the dominant component of IC flux in the jet, and the electron density can be much lower than in a pure IC/CMB jet. However, it means that our model can only exist in a narrow region of parameter space. While the model shows how the X-ray jet could in principle exist as an IC relic, it raises the question of why the electron spectrum in the jet would have exactly this shape. 

An additional complication of the relic jet model is the existence of jet knots. In modelling the jet pressures, we have assumed that the X-ray emission per unit volume is uniform throughout the jet.  However, the deep \textit{Chandra} exposure of the system reveals that there are several bright jet knots, most notably on the eastern side.  We estimated the contrast of these brightness variations. We took the average counts per pixel in regions J1-J3, and subtracted the average counts per pixel from L1-L3.  We find 40 counts per pixel on average in L1-L3, with a minimum of 5 and a maximum of 80 counts in a pixel. We repeated the procedure on the western side and found a pixel average of 15 counts, with minimum 0 and maximum 38 counts. This shows that there are significant brightness variations in both jets. This implies large variations in pressure along the jet.

To test the pressure variations in our model, we took one of the brightest parts of the eastern jet, in the middle of region J2. We defined a square region with the same width and orientation as the J2 jet region, and with a length of 4.5\arcsec centered on the brightest part. We then determined the background-subtracted average counts per pixel in this region to be 67 counts, or 1.67 times the  average counts per pixel over the entire jet.  We then ran \textit{synch} for this region, using a flux density of 1.67 times higher than  the flux density found in the eastern jet. The radio flux was scaled to the volume of the emitting region, but was not increased by a factor of 1.67, as the radio flux from the X-ray jet region is an upper limit. Modelling the jet knot in this way, we find pressures of $\sim 13\times 10^{-10}$ erg cm$^{-3}$ for $\gamma_{min}=1$ or $\sim 6 \times 10^{-10}$ erg cm$^{-3}$ for $\gamma_{\rm min}=10$.  This narrows the range of parameters pressure balance with the surrounding lobe somewhat, especially given the fact that $\gamma_{\rm min}$ is likely to be low in the relic jet. 

Unfortunately, the assumptions that we have made in trying to model the jet-like features are just too uncertain to be able to draw any definite conclusions about how whether the IC relic jet model is sustainable in Cyg A. While we find slightly higher pressures in the eastern jet compared to the lobe, both $\kappa$ and $\gamma_{\rm min}$ are unknown quantities that can greatly influence the pressure in the relic jet. 

Although  we can not rule out an IC relic jet model based on the X-ray and radio spectra, it is unclear how these knots of bright emission in the X-ray jets could be maintained in the IC relic jet model, where the radio jets have expanded by a factor of a few from their original, rather narrow size. If the knots originate from the radio jet, and if the X-ray jet has reached pressure equilibrium with the surrounding lobe, the knots should have been smoothed out in the process. 

\section{Conclusion}
\label{sec:conclusion}

Deep \textit{Chandra} observations of the X-ray jets and lobes of Cyg A have allowed us to analyse the emission from these features in detail. In the lobes, we have used two different tests to compare thermal and non-thermal models. In both lobes, we find that spectral fits strongly prefer a model with a non-thermal emission component for the lobe emission. In the X-ray jets, we used the thermal and non-thermal components of the ICM and lobe emission, and compared between a model with an additional thermal or additional non-thermal component for the jet emission. We find that the model with a non-thermal component for the jet is moderately to strongly preferred. 

MCMC sampling of the non-thermal lobe and jet models has given us constraints on the flux and photon indices of the jets and lobes.  For the eastern lobe and jet, we find $1$ keV flux densities of $71_{-10}^{+10}$ nJy and $24_{-4}^{+4}$ nJy, and photon indices of $1.72_{-0.03}^{+0.03}$ and $1.64_{-0.04}^{+0.04}$ respectively. For the western lobe and jet, we find flux densities of $50_{-13}^{+12}$ nJy and $13_{-5}^{+5}$ nJy, and photon indices of $1.97_{-0.10}^{+0.23}$ and $1.86_{-0.12}^{+0.18}$ respectively.

For each lobe, we used broken power laws with an age break to model the electron energy distributions. A comparison with the rim pressures from \cite{Snios2018} shows that a significant population of non-radiating particles is required to account for the total pressure of the eastern lobe. We also find a magnetic field of around $40$ $\mu$G, a factor 2 lower than the equipartition value of $73 - 95 $ $\mu$G . This ratio of $B/B_{eq}$ agrees well with a sample of other FRII radio galaxies \citep{Ineson2017}. 

However, in the western lobe no population of non-radiating particles is required and the low energy cutoff of the electron distribution needs to be raised to obtain a pressure consistent with the rim pressure. This discrepancy is a consequence of the difference in photon index between the two lobes, and suggests that the true electron distributions may be more complex than a single broken power law. A possible cause for the discrepancy is that the SSC component of the spectrum in the western lobe could have a turnover at lower energies than in the eastern lobe, perhaps because of the difference in size.  A spectral turnover below a few keV would yield a higher photon index in the $0.5 - 7.0$ keV energy range. The data are slightly suggestive of a turnover in this range, with a lower photon index between $0.5 - 2.0$ keV than at $2.0 - 7.0$ keV, although the constraints are not strong. A further complicating factor is the fact that the photon index likely isn't constant throughout the lobe. The correlation plots in Appendix \ref{appendix:corrs} show that the inner regions on the western side  (L4 and L5) contribute more to the total flux than the inner regions on the eastern side (L2 and L3). Because these regions presumably have a higher photon index, the average photon index for the western lobe will be higher as well. 

Regardless of the cause, if the photon index in the western lobe is overestimated, then the pressures on that side are overestimated as well. We consider the photon index in the eastern lobe more likely to be accurate because it agrees well with values in the literature of the average radio spectral index in the lobes. 

We modeled the X-ray jets according to the IC relic jet model from \cite{SB2008}. Similar to the lobes, we used the X-ray and radio data to constrain the spectra and model the electron distributions. On the 150 MHz LOFAR data, a weak brightness enhancement is seen at roughly the location of the relic jet, although imperfectly aligned. This could indicate the presence of the IC relic jet in radio wavelengths, but the emission is too weak to be able to constrain the radio synchrotron spectrum. Therefore, assumptions have to be made about the normalisation of the radio frequency and the break frequency. Moreover, $\gamma_{\rm min}$ and $\kappa$ are poorly constrained, making the modelling uncertain. 

We find a higher median pressure in the eastern jet compared to the lobe, but still within the errors. This suggests that an IC relic jet could be relatively close to pressure balance with the surrounding lobe. However, the IC relic jet model as we have modeled it can only exist in the narrow region of parameter space.  The magnetic field needs to be lower than in the lobe, so that the jet is not brighter than the lobe in the radio. At the same time, if the IC flux is dominated by SSC, the magnetic field needs to be high enough to produce synchrotron photons that can be scattered to the kev energy range.

An additional problem is that the model does not take into account the jet knots of increased brightness. If these knots originate from the radio jet, we would expect them to have been smoothed out in the process of adiabatic expansion if the current jet is close to achieving pressure balance with the lobe. Better constraints on the synchrotron spectrum, and more detailed modelling to allow for variations in flux and photon index along the jet axis, will be needed to shed more light on the nature of the jet-like X-ray features in Cyg A.

\section*{Acknowledgements}

We would like to thank the anonymous referee for their useful comments on the draft version of this paper. Support for this work was provided by the National Aeronautics and Space Administration through Chandra Award Number GO5-16117A issued by the Chandra X-ray Observatory Center, which is operated by the Smithsonian Astrophysical Observatory for and on behalf of the National Aeronautics Space Administration under contract NAS8-03060. PEJN was supported in part by NASA contract NAS8-03060. MJH acknowledges support from the UK Science and Technology Facilities Council [ST/M001008/1].




\bibliographystyle{mnras}
\bibliography{P1_papers} 



\appendix

\section{Statistical tools}
\label{appendix:stats}
\subsection{Bayesian inference}
 
Bayesian inference is based on Bayes' theorem,
\begin{equation}
p(\Theta | \mathbf{D},M)  = \frac{p( \mathbf{D} | \Theta, M) p(\Theta | M)}{p( \mathbf{D}| M)}.
\label{eq:Bayes}
\end{equation}
We compare a data set, $\mathbf{D}$, with a model $M$, which contains a set of parameters $\Theta = \{\Theta_{1}, \Theta_{2}...... \Theta_{N}\}$. $p(\Theta | \mathbf{D},M)$ is referred to as the \textit{posterior}.  $ p(\mathbf{D} | \Theta, M)$  is the probability of the data given parameters $\Theta$ and model $M$, and is referred to as the \textit{likelihood}.  $p(\Theta | M)$ encodes our prior knowledge of the system and is called the \textit{prior}. The normalising constant, $p(\mathbf{D}|M)$ is called the \textit{marginal likelihood}, and is the product of the likelihood and the prior integrated over the entire parameter space. The marginal likelihood is an important quantity in model selection and can be used to calculate the relative odds of two different models. However, calculating the marginal likelihood is computationally expensive for a model with a large number of parameters. Therefore we use unnormalised posteriors.

For computational convenience, we take the natural log of Eq. \ref{eq:Bayes}.  This allows us to sum the log terms of the equation. We will refer to the log terms as loglikelihood, logprior and logposterior.

The data set $\mathbf{D}$ consists of data points $D_{ij}$ for spectrum $i$  and spectral bin $j$, with a total of N spectra and J spectral bins per spectrum. The likelihood of the data, given  $M$ with parameters $\Theta$,  is the Poisson distribution multiplied over each bin of the data set,
\begin{equation}
L(\Theta) = \prod_{i=0}^N \prod_{j=0}^J \Big\{\frac{e^{-m(\Theta)} m(\Theta)^{D_{ij}}}{D_{ij}!} \Big\}
\end{equation}
By taking the log of this equation, we obtain the Poisson loglikelihood,
\begin{equation}
\log(L(\Theta)) = \sum_{i=0}^{N}  \sum_{j=0}^{J} \Big\{ - m(\Theta) + D_{ij}\log(m(\Theta)) - \log(D_{ij}!) \Big\}
\end{equation}
A \textit{Maximum Likelihood Estimation}, or MLE, estimates the most likely parameters $\hat{\Theta}$ by finding the parameters that maximise the likelihood. The Bayesian equivalent to a MLE is called a \textit{Maximum A Posteriori estimation}, or MAP. It is the set of parameters that maximises the posterior,
\begin{equation}
\hat{\Theta}_{\rm MAP} = \argmax_{\Theta} {( p( d | \Theta, M) p(\Theta | M))}
\end{equation}
The MAP is a useful tool to find the area of maximum likelihood.  This is useful, both for the model selection tests as well as to provide starting parameters for MCMC.

\subsection{Markov Chain Monte Carlo sampling}
\label{sec:MCMC}

In problems with large numbers of parameters, the posterior can become a highly complicated function. This is why we use Markov Chain Monte Carlo (MCMC) to sample the posterior. MCMC generates random samples of parameters by moving through the parameter space with Markov chains. Provided that these chains converge in a certain region of parameter space, the sampled parameter sets will  the posterior distribution of the parameters.

We used the Python module \textit{emcee} of \cite{Foreman-Mackey2013} which implements an affine invariant ensemble MCMC sampler based on \cite{Goodman2010}. The initial parameters for each of the chains are generated by a normal distribution centred on the MAP estimate for that parameter set.

The output of MCMC is a large set of sampled parameters for the given model.  Because we assume that the MCMC sampling maps the posterior function, these samples show us the values each parameter can have, and the likelihoods of these values. The MCMC sampled distribution of a parameter is therefore referred to as a a \textit{posterior distribution}. 

\subsection{Model selection methods}
\label{sec:mselection}

\subsubsection{Akaike/Bayesian information criterion}

\begin{table}
\centering
\caption{The Jeffreys scale. }
\begin{tabular}{cccc}
\hline \hline
$\Delta AIC/\Delta BIC$ & Odds & Probability & Strength of evidence \\ \hline
<1.0 & $\lesssim$  3:1 & 0.750 & Inconclusive \\ 
1.0 & $\sim$  3:1 & 0.750 & Positive evidence \\
2.5 & $\sim$  12:1 & 0.923 & Moderate evidence \\
5.0 & $\sim$  150:1 & 0.993 & Strong evidence \\ \hline

\label{tab:Jeffreys} 

\end{tabular}
\end{table}

The most important quantity in Bayesian model selection is the Bayes factor, also called the \textit{evidence}. The evidence is the ratio of marginal likelihoods of the models. If we consider Bayes' theorem for a model $M_{i}$, it is written as
\begin{equation}
p(M_i | \mathbf{D}) \propto p(\mathbf{D} | M_i) p(M_i) .
\end{equation}
If we now assume that the priors for both models are equal: $p(M_0) = p(M_1) = 0.5 $, we can write
\begin{equation}
B_{01} = \frac{p(M_1 | \mathbf{D} )}{p(M_0 | \mathbf{D})} = \frac{p(\mathbf{D} | M_1 )}{p(\mathbf{D} | M_0)} .
\end{equation}
Therefore, the ratio of marginal likelihoods of two models is equal to the ratio of the likelihood of the models. Because the marginal likelihood is difficult and computationally expensive to compute, we often use approximations. The Akaike Information Criterion (AIC, eq. \ref{AICeq}) \citep{Akaike1974},  and the Bayesian Information Criterion (BIC, eq. \ref{BICeq}) \citep{Schwarz1978},  are two such approximations that approach the marginal likelihood under certain circumstances.   

AIC and BIC include two terms: one for the \textit{complexity} of the model and one for the \textit{likelihood}. These terms have opposite signs. A model with more parameters will always be able to give a higher likelihood in a fit, but adding 'unnecessary' extra parameters will result in overfitting.  Therefore the best model is the one that has the highest likelihood while being the least complex. 

The AIC is defined as 
\begin{equation}
\label{AICeq}
{\rm AIC} = 2k - 2\ln{\hat{L}}
\end{equation}
and the BIC as 
\begin{equation}
\label{BICeq}
{\rm BIC} = \ln(n)k - 2\ln{\hat{L}},
\end{equation}
where \textit{k} is the number of free parameters in the model, $\hat{L}$ is the maximum likelihood of the model fit to the data, and \textit{n}  the number of independent data points. Eqs \ref{AICeq} and \ref{BICeq} differ only in the first term. Because most of the time $\ln(n)k > 2k$, the BIC prefers less complex models more strongly.

The model with the lowest AIC or BIC is considered the most likely model. The relative odds of two models can be calculated from the difference in AIC/BIC values. The scale of relative odds between two models is known as the Jeffreys Scale \citep{Kass1995, Jeffreys1961, Trotta2007}. An example of $\Delta AIC/\Delta BIC$ and corresponding relative odds is shown in Table \ref{tab:Jeffreys} \citep[taken from][]{Trotta2007}.
 
The AIC assumes a large sample size. For smaller sample sizes, a correction needs to be applied. This is called the corrected AIC or, AIC$_{\rm C}$ \citep{Burnham2002} 
\begin{equation}
AIC_C = 2k - 2\ln \hat{L} + \frac{2(k+1)(k+2)}{n-k-2}.
\label{AICCeq}
\end{equation}
While AIC and BIC are widely-used in model selection, they also have certain disadvantages. They depend only on the maximum likelihood, which means the prior information is not considered. Secondly, a model with degenerate, and therefore unconstrained, parameters will be overly penalised by AIC and BIC : the model complexity increases with little gain to the goodness-of-fit. The number of free parameters is therefore not necessarily a good indication of model complexity \citep{Liddle2007}.

\subsubsection{Likelihood-ratio test}

The likelihood ratio test is a form of hypothesis testing for nested models. We form our null hypothesis, $H_{0}: M = M_{0}$, and the alternative hypothesis, $H_{1}: M = M_{1}$. We define the Likelihood-ratio test statistic, $T_{\rm LRT}$, as:

\begin{equation}
T_{\rm LRT} = \ln(\frac{p(\mathbf{D} | \hat{\Theta _{1}}, M_{1})}{p(\mathbf{D} | \hat{\Theta _{0}}, M_{0})})
\end{equation} 

We find the likelihood ratio of models $M_{1}$ and $M_{0}$ by determining the likelihood which corresponds to the MAP estimate for each model.  Assuming $H_{0}$, we want to find out where the likelihood ratio lies on the distribution of possible likelihood ratios between $M_{1}$ and $M_{0}$. If the observed ratio lies in the tail of the likelihood ratio distribution, it would be very unlikely to occur by chance under the null hypothesis. By comparing the $T_{\rm LRT}$ with the distribution of likelihood ratios, we can decide whether we have cause to reject $M_{0}$ or not. 

The downside of hypothesis testing is that the p-value tells us the probability that we can reject $M_{0}$. It does not say anything  about the likelihood of $M_1$. 

To obtain a distribution of possible likelihood ratios, we need predictive data assuming the null hypothesis. We MCMC sampled model $M_{0}$ for the eastern and western lobe.  The MCMC sampling was done in as described in section \ref{sec:MCMC}. We then picked a random parameter set from the sample and generated a fake spectrum with the \texttt{Sherpa} tool \textit{fake\_pha}. This fake spectrum is effectively a simulated future observation of the data, assuming $H_{0}$. We again find the likelihoods corresponding to the MAP estimates for model $M_{0}$ and $M_{1}$ to the simulated spectrum and calculate the likelihood ratio. By drawing repeated samples, generating fake data, and calculating the likelihood ratio, we obtain a likelihood ratio distribution to which we can compare the likelihood ratio of the real data. From this we can calculate the p-value.

\section{Correlation plots}
\label{appendix:corrs}
\begin{figure*}
   \includegraphics[width=0.9\textwidth]{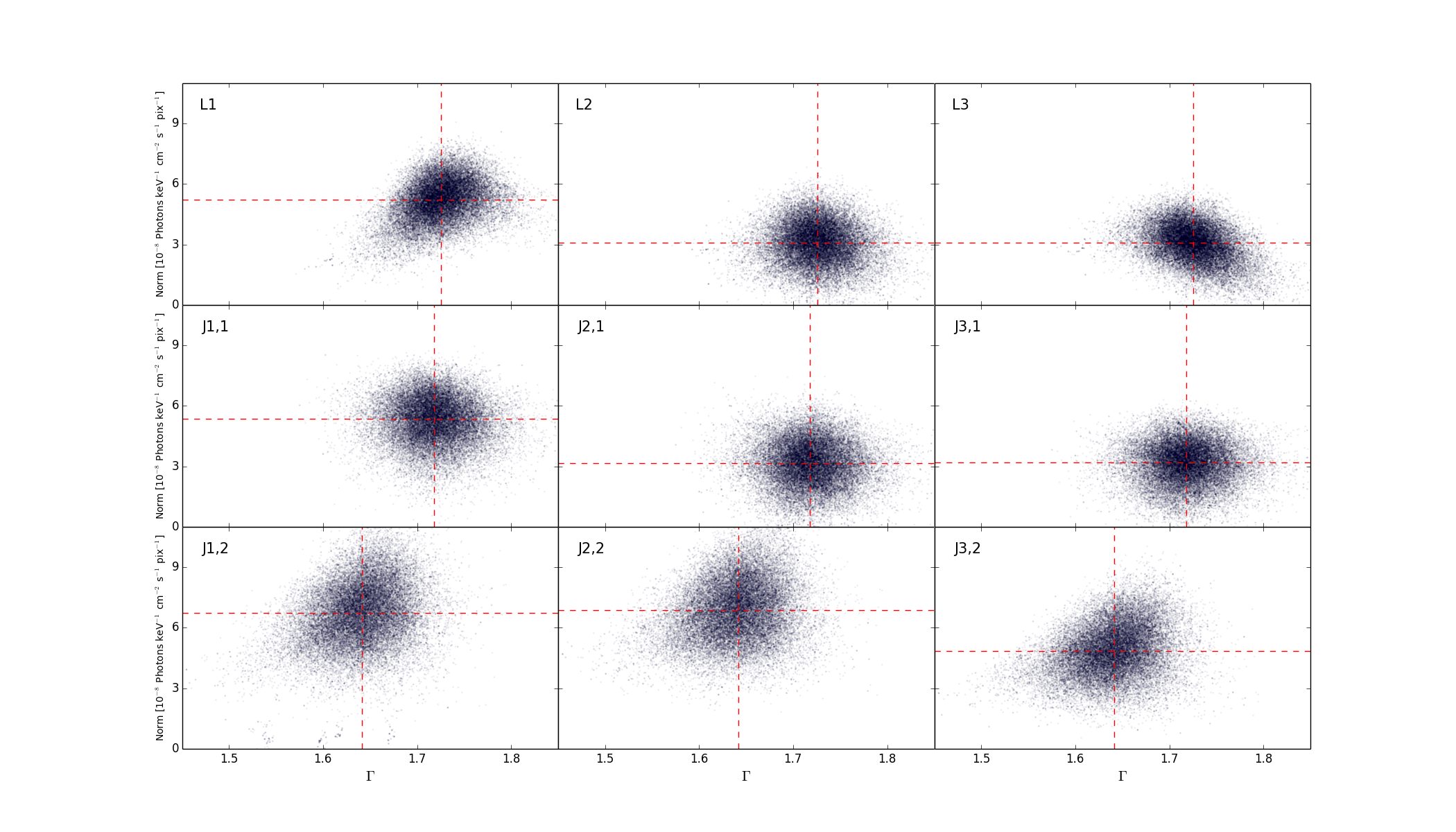}
   \includegraphics[width=0.9\textwidth]{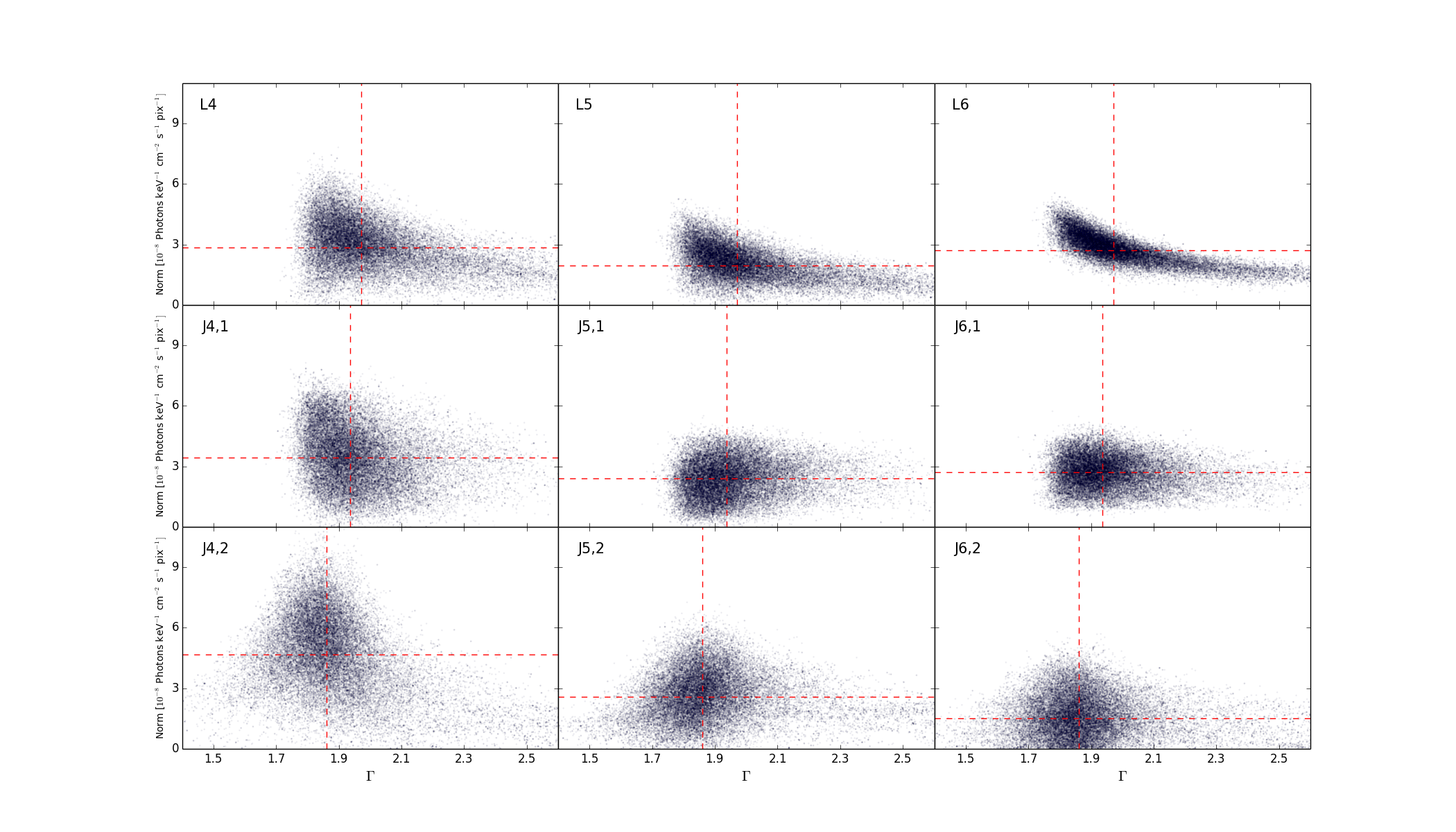}
  \caption{Top:  Scatter plots that show the correlation between $\Gamma$ and surface brightness for each non-thermal emission component in each region of the eastern lobe. The surface brightness is expressed as the spectral normalisation of the APEC model divided by the pixel area of the region. The dashed horizontal and vertical red lines indicate the median normalisations and photon indices respectively. The labels in the top left indicate the region. In the jet regions the emission component is additionally specified. I.e. J1,1 indicates the non-thermal emission from the lobe in region J1. Bottom: As above, for the western lobe.} 
    \label{fig:corrplots}

\end{figure*}


\bsp	
\label{lastpage}
\end{document}